\pdfoutput=1 

\documentclass[%
,aps%
 ,preprint,tightenlines%
 ,secnumarabic%
 ,nofootinbib
,amssymb, amsmath,nobibnotes, prl, floatfix]{revtex4}
\usepackage{docs}%
\usepackage{bm}%
\usepackage{hyperref}%
\expandafter\ifx\csname package@font\endcsname\relax\else
 \expandafter\expandafter
 \expandafter\usepackage
 \expandafter\expandafter
 \expandafter{\csname package@font\endcsname}%
\fi

\usepackage[italian, english]{babel}
\usepackage{amsmath}                                      
\usepackage{amsthm}

\usepackage{cancel}

\usepackage{wrapfig}
\usepackage{amssymb}
\usepackage{comment}
\usepackage{latexsym}
\usepackage{mathtools}
\usepackage{dsfont}
\usepackage{mathrsfs}
\usepackage{hyperref}
\usepackage{float}
\usepackage{graphicx}

\usepackage{framed}
\usepackage{amsfonts}

\usepackage{pgf}
\usepackage{physics}

\usepackage{tikz-feynman}

\newcommand{\beq}{\begin{equation}}
\newcommand{\eeq}{\end{equation}}
\newcommand{\beqa}{\begin{eqnarray}}
\newcommand{\eeqa}{\end{eqnarray}}
\newcommand{\beqar}{\begin{eqnarray*}}
\newcommand{\eeqar}{\end{eqnarray*}}

\def\Tr           {\mbox{\rm Tr}\,}
\def\STr          {\mbox{\rm STr}\,}

\def\fsH    {H\!\!\!\!/\,}

\def\sst#1{{\scriptscriptstyle #1}}
\def\0{{\sst{(0)}}}
\def\1{{\sst{(1)}}}
\def\2{{\sst{(2)}}}
\def\3{{\sst{(3)}}}
\def\4{{\sst{(4)}}}
\def\5{{\sst{(5)}}}
\def\6{{\sst{(6)}}}
\def\7{{\sst{(7)}}}
\def\8{{\sst{(8)}}}

\usepackage[utf8]{inputenc}

\usepackage{amsmath}
\usepackage{amssymb}
\usepackage{physics}
\usepackage[inline]{asymptote}

\newcommand{\A}{\mathcal{A}}
\newcommand{\Hp}{{\mathcal{H}^+}}
\newcommand{\Beta}{\operatorname{B}}
\newcommand{\Fxz}[5]{F^{#1,#2}_{#3,#4,#5}}

\usepackage{xcolor}

\newenvironment{eqaed}
    {\begin{equation}
    \begin{aligned}
    }
    { 
    \end{aligned}
    \end{equation}
    }

\makeatletter
\preto\maketitle{%
  \begingroup\lccode`~=`,
  \lowercase{\endgroup
  \let\saved@breqn@active@comma~% save breqn active comma
  \let~}\active@comma % set the active comma to what revtex4-1 wants
}
\appto\maketitle{%
  \begingroup\lccode`~=`,
  \lowercase{\endgroup
  \let~}\saved@breqn@active@comma % undo the change
}
\makeatother
\numberwithin{equation}{section}

\begin{document}

% title page data

\title{On All-Order Higher-Point $\text{D}p$--$\overline{\text{D}p}$ Effective Actions}

\author{Riccardo Antonelli}
\email{riccardo.antonelli@sns.it}

\author{Ivano Basile}
\email{ivano.basile@sns.it}
\affiliation{{Scuola Normale Superiore and I.N.F.N.\\Piazza dei Cavalieri 7, 56126, Pisa, Italy}}

\author{Ehsan Hatefi}
\email{ehsan.hatefi@sns.it}
\affiliation{{Scuola Normale Superiore and I.N.F.N.\\Piazza dei Cavalieri 7, 56126, Pisa, Italy}}
\affiliation{{Mathematical Institute\\ Faculty of Mathematics\\Charles University, P-18675, CR\\}} 

\date{\today}

\begin{abstract}
\noindent In this paper we derive a class of contributions to all orders in $\alpha'$ to the effective action of D-brane-anti D-brane systems in Type II String Theory, considering an amplitude involving a closed-string Ramond-Ramond state and four open-string states: two tachyons, a scalar and a gauge boson. This type of amplitude arises in both Type IIA and Type IIB strings, and reveals a number of new effective couplings. Furthermore, we derive a series representation for the result that goes beyond a factorized limit that was recently studied, and which is expected to apply to more general six-point amplitudes of superstrings, including external fermions.
\end{abstract}

% title page and toc

\maketitle
\tableofcontents

% MAIN DOC
% set baseline skip for the rest
\baselineskip=18pt

\section{Introduction}\label{sec:intro}

%structure of ST corrections and new couplings
%susy breaking
%time-dependent bgs
%amplitude formalism is great
%d-branes and open strings
%holographic applications
%open tachyons instability applications
%previous works, non-BPS brane decay, tachyon condensation of ddbar, all-order stuff
%inflation and kklt applications
%no dualities
%tachyon dynamics polchinski
%amplitudes fix coefficients and stuff
%large volume scenario
%WZ, DBI, myers
%paper outline
%CITAZIONI DA METTERE

String Theory provides an ultraviolet completion to gravity that brings along systematic corrections to low-energy interactions.
These can be quantum corrections, which are weighted by the string coupling $g_s$, or world-sheet curvature corrections, which are controlled by the ``Regge slope'' $\alpha'$.
A deeper understanding of the resulting phenomena cannot forego a better control of all-order, and eventually non-perturbative, effects.
This paper is devoted to the study all-order $\alpha'$ corrections in a class of mixed amplitudes that comprise open strings~\footnote{For in-depth discussions of open strings, see~\cite{orientifold}.} and one closed string. More specifically, we focus on the \textit{D-brane-anti D-brane} ($\text{D}p$-$\overline{\text{D}p}$) system~\footnote{For earlier detailed treatments, see~\cite{Hatefi:2017ags,Michel:2014lva,Bergman:1998xv}.}, an important ingredient in the breaking of supersymmetry where the generic emergence of tachyonic modes signals a tendency towards the annihilation of D-branes with anti D-branes.
We extract corrections to the low-energy effective action from an amplitude involving a closed-string Ramond-Ramond (RR) state, two open-string tachyons, an open-string massless scalar and an open-string gauge boson, which entails some subtleties. One can compute amplitudes of this type using Conformal Field Theory (CFT) techniques~\cite{Friedan:1985ge}, and the results generalize the original one from Veneziano~\cite{Veneziano:1968yb}. At low energies, the last two states correspond to ``diagonal'' massless fields living on the world-volume of the brane stacks, while the tachyon arises from a mixing mode, and is thus ``off-diagonal''.

Corrections of this type unveil generically new couplings~\footnote{Similar corrections can also be found in the case of S-branes~\cite{s-branes}.}~\cite{Hatefi:2012ve,Kennedy:1999nn}, and gaining a better control of them can also shed some light on the key issue of supersymmetry breaking. Important open questions concern non-perturbative dualities, whose role in this context was first analyzed in~\cite{Blum:1997cs}. In addition, the emergence of tachyons accompanies new phenomena, absent in the supersymmetric case, which include their condensation~\cite{Sen:2002in,Sen:2002an,Sen:1998sm} and the decay of non-BPS D-branes~\cite{Lambert:2003zr,Sen:2004nf,Sen:1999mg,Sen:1999md,Bergshoeff:2000dq}.

Moreover, all-order $\alpha'$ effects can also be important in the study of cosmological backgrounds, and in particular inflationary models~\cite{Dvali:1998pa,Choudhury:2003vr,Kachru:2003sx}, where regions of strong curvature are generically present close to the initial singularity. Possible applications of this type of analysis to non-gravitational gauge theories also include (top-down) holographic models~\cite{Maldacena:1997re,Witten:1998qj} with broken supersymmetry, as for instance the Sakai-Sugimoto model~\cite{Sakai:2004cn,Sakai:2005yt,Okuda:2002yd} of holographic QCD~\cite{Casero:2007ae,Dhar:2007bz}. Typically, going beyond the (super)gravity approximation translates into wider ranges of validity for the 't Hooft coupling.
Finally, let us mention metastable de Sitter vacua~\cite{Polchinski:2015bea,deAlwis:2013gka} as another possible avenue of investigation. $\text{D}$--$\overline{\text{D}}$ systems may be also held responsible for the uplift term in the KKLT construction~\cite{Kachru:2003aw} and its extensions. Notice that the combined effects of anti-branes and orientifolds give rise to the same type of uplift in Brane Supersymmetry Breaking~\cite{bsb}, where however tachyons are not present.

Perturbative scattering amplitudes involve structures that are largely constrained, which makes it possible to reconstruct them, up to an overall coefficient, via Effective Field Theory (EFT) analyses of poles and symmetries in various channels.
On the other hand, \mbox{\textit{ab-initio}} computations of the corresponding string amplitudes have the virtue of fixing the $\alpha'$ expansion, and thus the EFT couplings, to all orders. This procedure has been followed systematically in the literature~\cite{Bjerrum-Bohr:2014qwa,Hatefi:2014lva,Barreiro:2013dpa,Hatefi:2014saa,Chandia:2003sh,Hatefi:2016yhb}, yielding generalizations of the standard Dirac-Born-Infeld (DBI), Myers~\cite{Myers:1999ps} and Wess-Zumino (WZ) terms~\footnote{Other techniques have been employed to determine effective actions, including supersymmetric localization~\cite{Hashimoto:2015iha}.}~\cite{Hatefi:2010ik}. The crucial interplay between D-branes and RR potentials has been thoroughly studied in the literature~\cite{Polchinski:1995mt}, but their role is yet to be fully dissected.

The paper is organized as follows. In Section~\ref{sec:lower} we summarize a number of results that were obtained in earlier works, which provides a starting point for our considerations. Specifically, we discuss all-order $\alpha'$ corrections to couplings that originate from amplitudes a single closed-string mode and less than four open-string ones. We focus on RR closed-string states, and thus on couplings related to WZ terms.
In Section~\ref{sec:higher} we present the new result on the five-string scattering amplitude under consideration. From the point of view of the (open) world-sheet CFT, this is akin to a six-point correlator, whose vertex operators are to be integrated accordingly. The result is organized in terms of a class of integrals, whose detailed study is postponed to Section~\ref{sec:beyond}. In Section~\ref{sec:factorized_limit} we briefly discuss a factorized limit in which the integrals can be evaluated in closed form. In sections~\ref{sec:vector_bulk_sing} and~\ref{sec:scalar_bulk_sing}, after a proper discussion of bulk singularities, selection rules~\cite{Hatefi:2013yxa} and EFT methods, we recover the result from the low-energy perspective.
In Section~\ref{sec:beyond} we go beyond the factorized limit of Section~\ref{sec:factorized_limit}, obtaining an expansion for the integrals, whose leading term reproduces it. While the resulting expansion entails a limit of series with decreasing convergence rate, we show that the procedure is well-defined. We conclude in Section~\ref{sec:conclusions} with a summary of our results, and some remarks concerning their potential implications.

The paper contains two appendices. In Appendix~\ref{appendix:qandns} we provide explicit expressions for the integrals involved in the factorized limit of Section~\ref{sec:factorized_limit}.
In Appendix~\ref{appendix:z_integral} we derive a closed-form expression for the reduced integrals involved in the factorized limit.
    
\section{All-Order Corrections to Lower-Point Interactions}\label{sec:lower}

In this section we summarize a number of results concerning scattering amplitudes for less than five strings~\footnote{We remark that, in open string scattering amplitudes, each closed-string vertex counts is tantamount to two open-string ones. Hence, five-string scattering processes involving one closed string involve six-point world-sheet correlators.}. For more detailed discussions on the material presented in this section, see~\cite{Hatefi:2012cp,Hatefi:2017ags}. To begin with, let us recall the expressions for the relevant vertex operators,
\begin{eqaed}\label{eq:vertex_operators}
    V_{T}^{(0)}(x) & = \alpha' \, ik_1 \cdot \psi(x) \, e^{\alpha' \, ik_1 \cdot X(x)} \, \lambda \otimes \sigma_1 \, , \\
    V_A^{(-1)}(x) & = e^{-\phi(x)} \, \xi_2^a \psi_a(x) \, e^{\alpha' \, i k \cdot X(x)} \, \lambda \otimes \sigma_3 \, , \\
    V_{\phi}^{(0)}(x) & = \xi_1^i \left( \partial X_i(x) + \alpha' \, i q \cdot \psi(x) \, \psi_i(x) \right) \, e^{\alpha' \, i q \cdot X(x)} \, \lambda \otimes I \, ,  \\
    V_{C}^{(-\frac{1}{2},-\frac{1}{2})}(z,\bar{z}) & = \left(P_- \fsH_{(n)} M_p \right)^{\alpha \beta} \, e^{-\frac{\phi(z)}{2}} \, S_\alpha(z) \, e^{\frac{\alpha'}{2} \, i p \cdot X(z)} \, e^{-\frac{\phi(\bar{z})}{2}} \, S_\beta(\bar{z}) \, e^{\frac{\alpha'}{2} \, i p \cdot D \cdot X(\bar{z})} \, \otimes \sigma_3 \, ,
\end{eqaed}
including the corresponding $U(N)$ Chan-Paton (CP) factors~\cite{Paton:1969je,Schwarz:1974vk,Marcus:1982fr} $\lambda$ and the Pauli matrices $\sigma_i$. The latter account for where the endpoints of the open strings terminate and for the sign of the RR coupling. Combining the $\sigma_i$ are combined with the $\lambda$ in this fashion reconstructs the full $U(N) \times U(N)$ CP factors. The numbers in superscripts denote \textit{superghost charge}, while $\xi_{1,2}$ are the polarizations associated to the states. We work in a picture where the vertex operator of the world-volume scalar has vanishing superghost charge~\footnote{This choice has the virtue of producing consistent results, despite some subtleties with picture changing in the presence of world-volume scalars that are discussed in~\cite{Hatefi:2015jpa}.}. For the expressions of vertex operators in different pictures, see~\cite{Hatefi:2015okf}. Finally, the $S_\alpha(z)$ are \textit{spin operators}, which build RR states.

We denote ten-dimensional spacetime indices by $\mu \, , \nu = 0 \, , 1 \, , \, \dots \, , 9$, and world-volume and transverse directions by $a \, , b \, , c = 0 \, , 1 \, , \, \dots \, , p$ and $i \, , j = p+1 \, , \, \dots \, , 9$ respectively.

Furthermore, we introduce momenta $q \, , p \, , k \, , k_1$, whose on-shell conditions read
\begin{equation}\label{eq:on_shell_cond}
    q^2 = p^2 = k^2 = 0 \, , \qquad k_1^2 = \frac{1}{4} \, , \qquad q \cdot \xi_{1} = k \cdot \xi_{1,2} = 0 \, ,
\end{equation}
the projector
\begin{equation}\label{eq:projector}
    P_- = \frac{1}{2}\left(1 - \gamma^{11} \right)
\end{equation}
and the field-strength of a generic Ramond-Ramond (RR) form $C_{(n-1)}$,
\begin{equation}\label{eq:form_slash}
    \fsH_{(n)} = \frac{a_n}{n!} \, H_{\mu_1 \dots \mu_n} \, \gamma^{\mu_1} \dots \gamma^{\mu_n} \, .
\end{equation}
For BPS branes the coefficient $a_n=i$ for even values of $n$, which pertain to Type IIA String Theory, and $a_n = 1$ for odd values of $n$, which pertain to Type IIB. Furthermore, we use the charge conjugation matrix $C$ to write 
\begin{equation}\label{eq:C_matrix_indices}
    \left(P_- \, \fsH_{(n)} \right)^{\alpha \beta} = C^{\alpha \delta} {(P_- \, \fsH_{(n)})_\delta}^{\beta} \, .
\end{equation}
The RR polarizations encoded in eq.~\eqref{eq:form_slash} build, in the low-energy EFT, the field strength associated to the RR form potential. Finally, we employ the \textit{doubling trick} to deal with the anti-holomorphic parts of the closed-string world-sheet fields, via the substitutions
\begin{eqaed}\label{eq:doubling_trick}
    \tilde{X}^\mu(\bar{z}) & \, \longrightarrow \, {D^\mu}_\nu \, X^\nu(\bar{z}) \, ,   \\
    \tilde{\psi}^\mu(\bar{z}) & \, \longrightarrow \, {D^\mu}_\nu \, \psi^\nu(\bar{z}) \, ,   \\
    \tilde{\phi}(\bar{z}) & \, \longrightarrow \, \phi(\bar{z}) \, ,   \\
    \tilde{S}_\alpha(\bar{z}) & \, \longrightarrow \, {{(M_p)}_\alpha}^\beta \, S_\beta(\bar{z}) \, ,
\end{eqaed}
where the matrices $D$ and $M_p$ are
\begin{equation}\label{doubling_D_Mp}
    D = \mqty(-\mathbf{1}_{9-p} & 0 \\ 0 & \mathbf{1}_{p+1}) \, , \qquad {{(M_p)}_\alpha}^\beta = \left\{\begin{array}{cc}\pm\frac{i}{(p+1)!} \, {{(\gamma^{i_1} \dots \gamma^{i_{p+1}})}_\alpha}^\beta \, \epsilon_{i_1 \dots i_{p+1}} \quad p \; \text{even} \\ \pm \frac{1}{(p+1)!} \, {{(\gamma^{i_1} \dots \gamma^{i_{p+1}} \, \gamma_{11})}_\alpha}^\beta \epsilon_{i_1 \dots i_{p+1}} \quad p \; \text{odd} \end{array}\right. \, .
\end{equation}
With these ingredients at hand, scattering amplitudes are determined by the free two-point functions of the $X^\mu \, , \psi^\mu \, , \phi$ fields,
\begin{eqaed}\label{eq:propagators}
    \langle X^\mu(z) \, X^\nu(w) \rangle & = - \, \frac{\alpha'}{2} \, \eta^{\mu \nu} \, \log(z-w) \, ,   \\
    \langle \psi^\mu(z) \, \psi^\nu(w) \rangle & = - \, \frac{\alpha'}{2} \, \eta^{\mu \nu} \, (z-w)^{-1} \, ,  \\
    \langle \phi(z) \, \phi(w) \rangle & = - \, \log(z-w) \, .
\end{eqaed}

While these expressions hold for the world-sheet CFT of strings in a flat background, scattering amplitudes can be used to reconstruct the effective action in complete generality, taking into account diffeomorphism and gauge invariance. In addition to the propagators, one also needs correlations involving spin operators. The ``Wick-like rule'' discussed in~\cite{Liu:2001qa,Hatefi:2015ora} encompasses the relevant cases.

The resulting amplitudes can be expanded in Laurent series in the relevant Mandelstam variables, whose coefficients are instrumental in the matching with EFT. This procedure determines which terms are allowed in the effective action, while perturbative string amplitudes provide their coefficients to \textit{all orders} in the $\alpha'$ expansion. For instance, as discussed in~\cite{Hatefi:2016yhb}, $C \, T \, T$ scattering amplitudes generate in this fashion all-order corrections of the form
\begin{equation}\label{eq:S_C-DT-DT}
    i \mu_p \, (2 \pi \alpha')^2 \int_{\Sigma_{(p+1)}} \!\!\!\!\!\!\! C_{(p-1)} \wedge \Tr\! \left[\sum_{m=-1}^\infty c_m \, (\alpha')^{m+1} \left(D^b D_b \right)^{m+1} DT \wedge DT \right] \, ,
\end{equation}
consistently with an EFT analysis of world-volume singularities, while $C \, \phi \, A$ and $C \, \phi \, A \, A$ scattering reveal the presence of couplings of the form
\begin{equation}\label{eq:S_C-Dphi}
    \frac{\mu_p (2 \pi \alpha')}{(p+1)!} \int_{\Sigma_{(p+1)}} \!\!\!\!\!\!\! d^{p+1} \sigma \, \epsilon^{a_0 \dots a_p} \, {C^i}_{a_0 \dots a_{p-1}} \, D_{a_p} \phi_i
\end{equation}
and
\begin{equation}\label{eq:S_C-A-Dphi}
     \frac{\mu_p (2 \pi \alpha')^2}{(p-1)!} \int_{\Sigma_{(p+1)}} \!\!\!\!\!\!\! d^{p+1} \sigma \, \Tr\! \Big[ \epsilon^{a_0 \dots a_p} \, {C^i}_{a_0 \dots a_{p-3}} \, F_{a_{p-2} a_{p-1}} \, D_{a_p} \phi_i \Big] \, ,
\end{equation}
where $\phi_i$, not to be confused with the world-sheet bosonized ghost field $\phi$, are 
scalars that describe the embedding of branes in space-time. Here the transverse index $i$ is part of the $p \pm 1$ antisymmetric indices of the (components of the) RR forms $C_{(p \pm 1)}$. It is contracted with the embedding scalars $\phi_i$, so that the overall integrands have the appropriate rank.

One is thus led to the all-order corrections
\begin{eqaed}\label{eq:S_C-A-Dphi_corr}
    \frac{\mu_p (2 \pi \alpha')^2}{(p-1)!} \int_{\Sigma_{(p+1)}}\!\!\!\!\!\!\! d^{p+1} \sigma \, &\Tr\! \Bigg[ \epsilon^{a_0 \dots a_p} \, {C^i}_{a_0 \dots a_{p-3}} \, \sum_{n=-1}^\infty b_n (\alpha')^{n+1}   \\
    & D_{a_1} \dots D_{a_{n+1}} F_{a_{p-2} a_{p-1}} \, D^{a_1} \dots D^{a_{n+1}} D_{a_p} \phi_i \Bigg] \, ,
\end{eqaed} 
and similarly to the mixed Chern-Simons terms
\begin{eqaed}\label{eq:S_C-DRorT-Dphiorphi}
    & \frac{2i \, \beta \, \mu'_p}{p!} \, (2 \pi \alpha')^2 \int_{\Sigma_{(p+1)}} \!\!\!\!\!\!\! \partial_i C_{(p)} \wedge DT \, \phi^i \, ,   \\
    & \frac{2i \, \beta \, \mu'_p}{p!} \, (2 \pi \alpha')^2 \int_{\Sigma_{(p+1)}} \!\!\!\!\!\!\! {(C^i)}_{(p-1)} \wedge DT \wedge D \phi_i \, ,
\end{eqaed}
which were also revealed by direct computations. Here we use a mixed index/form notation~\footnote{The mixed index/form notation $(C^i)_{(p-1)}$ reads
\begin{equation}\label{eq:mixed_notation}
    (C^i)_{(p-1)} = \frac{1}{(p-1)!} \, {C^i}_{a_0 \dots a_{p-2}} \, dx^{a_0} \wedge \dots \wedge dx^{a_{p-2}} \, .
\end{equation}
}.
In addition, it was shown that the generalized Bianchi identities
\begin{eqaed}\label{eq:transverse_cond_H}
    & p_a \, H_{a_0 \dots a_{p-1}} \, \epsilon^{a_0 \dots a_{p-1} a} = 0 \, ,   \\
    & p^i \, \epsilon^{a_0 \dots a_p} \, H_{a_0 \dots a_p} + p_a \, \epsilon^{a_0 \dots a_{p-1} a} \, {H^i}_{a_0 \dots a_{p-1}} = 0 \, ,
\end{eqaed}
need to be imposed in order that the amplitude be picture-independent~\cite{Hatefi:2018qlm}.

The preceding expressions describe new couplings, with their all-order corrections, to be added to the Dirac-Born-Infeld (DBI) and Wess-Zumino (WZ) actions. They include tachyon contributions for the non-BPS combinations of branes~\cite{Hatefi:2017ags} under scrutiny. The modified DBI action, which involves a symmetric trace, reads
\begin{equation}\label{eq:S_DBI}
    S_{\text{DBI}} = - \, T_{p} \int_{\Sigma_{(p+1)}} \!\!\!\!\!\!\! d^{p+1} \sigma \, \Tr\! \left[V(\mathcal{T}) \, \sqrt{-\det \left(\eta_{ab} + 2 \pi \alpha' F_{ab} + 2 \pi \alpha' \, D_a \mathcal{T} D_b \mathcal{T} \right)} \right] \, .
\end{equation}
In order to describe the $\text{D}p$-$\overline{\text{D}p}$ system,
\begin{equation}\label{eq:F_12}
    F_{ab} = \mqty(F^{(1)}_{ab} & 0 \\ 0 & F^{(2)}_{ab}) \, ,
\end{equation}
where the $F^{(i)}_{ab} = \partial_a A^{(i)}_b - \partial_b A^{(i)}_a$ describe the individual factors of the $U(N) \times U(N)$ gauge group. On the other hand, the tachyon field is off-diagonal,
\begin{equation}\label{eq:T_12}
    \mathcal{T} = \mqty(0 & T \\ T^* & 0) \, ,
\end{equation}
and the corresponding instability reflects the mutual force between the D-branes and the anti D-branes. The gauge-covariant derivative reads, accordingly,
\begin{equation}\label{eq:DT_12}
    D_a \mathcal{T} = \mqty(0 & D_a T \\ (D_a T)^* & 0) \, ,
\end{equation}
with
\begin{equation}\label{eq:DT_def}
    D_a T = \partial_a T - i \, (A^{(i)} - A^{(2)}) \, T \, .
\end{equation}
Eqs.~\eqref{eq:T_12} and~\eqref{eq:DT_def} highlight the bi-fundamental nature of the tachyon field, which arises from open strings stretching between the two stacks.

The modified DBI action contains the tachyon potential
\begin{equation}\label{eq:VT}
    V(\mathcal{T}) = 1 + \pi \alpha' m^2 \, \abs{T}^2 + \frac{(\pi \alpha' m^2)^2}{2} \, \abs{T}^4 + \dots \, ,
\end{equation}
where $m^2 = -1/(2 \alpha') \equiv m^2_T$ is the tachyon mass. These first terms in the expansion are consistent with a Gaussian potential~\cite{Kutasov:2000aq}
\begin{equation}\label{eq:VT_exp}
	V(\mathcal{T}) = e^{\pi \alpha' m^2 \abs{T}^2} \, ,
\end{equation}
which would drive the tachyon field to condense at infinity. However, the couplings that arise from it cannot be directly compared to the results of Boundary String Field Theory (BSFT), since the low-energy expansion is performed around different values of the Mandelstam variables. While the couplings discussed in this paper only connect with the quadratic expansion of eq.~\eqref{eq:VT_exp}, the consistency of the quartic term was studied in~\cite{Hatefi:2017ags}. In addition to the new couplings discussed above, it was found that terms of the form~\footnote{The $\cdot$ symbol here denotes index contraction, wedge product and traced product over gauge degrees of freedom.} $F^{(1)} \cdot F^{(2)}$ and $D\phi^{(1)} \cdot D\phi^{(2)}$ are also present, and their coefficients were fixed by direct computations.

Let us now describe how the WZ action is modified in the $\text{D}p$-$\overline{\text{D}p}$ system. Leaving aside tachyons, it reads~\cite{Douglas:1995bn,Li:1995pq,Green:1996dd}
\begin{equation}\label{eq:S_WZ}
    S_{\text{WZ}} = \mu'_p \int_{\Sigma_{(p+1)}} \left[ C \wedge \left( e^{i \, 2 \pi \alpha' F^{(1)}} - e^{i \, 2 \pi \alpha' F^{(2)}} \right)\right]_{(p+1)}\, ,
\end{equation}
where the (formal) linear combination $C = \sum_n (-i)^{\frac{p-m+1}{2}} \, C_{(m)}$ is an element of the exterior algebra, and the $[\cdot]_{p+1}$ notation denotes projection onto the subspace of $(p+1)$-forms. The inclusion of tachyons rests on the superconnection described in~\cite{Quillen:1985vya,berl,Roepstorff:1998vh}, and modifies the expression according to
\begin{equation}\label{eq:S_WZ2}
    S_{\text{WZ}} = \mu'_p \int_{\Sigma_{(p+1)}}\left[ C \wedge \STr e^{i \, 2 \pi \alpha' \mathcal{F}} \right]_{(p+1)}  \, .
\end{equation}
The superconnection
\begin{equation}\label{eq:A_def}
    i \, \A = \mqty(i \, A^{(1)} & \beta \, T^* \\ \beta \, T & i \, A^{(2)})
\end{equation}
includes the tachyon field, accompanied by the normalization constant $\beta$ appropriate for the $\text{D}p$-$\overline{\text{D}p}$ system, which can be matched to a suitable amplitude~\cite{Hatefi:2016yhb}. Its curvature is defined, as usual, by
\begin{equation}\label{eq:F_def}
    \mathcal{F} = d\A - i \, \A \wedge \A \, ,
\end{equation}
which leads to
\begin{equation}\label{eq:iF}
    i \, \mathcal{F} = \mqty(i \, F^{(1)}  - \beta^2 \, \abs{T}^2 & \beta (DT)^* \\ \beta (DT) & i \, F^{(2)} - \beta^2 \, \abs{T}^2)
\end{equation}
in the present case. Here we used the form language, with $F^{(i)} = \frac{1}{2} \, F^{(i)}_{ab} \, dx^a \wedge dx^b$ and $DT = D_a T \, dx^a$, and some of the relevant structures that arise from the (formal) exponential of the superconnection in eq.~\eqref{eq:S_WZ2} evaluate to~\footnote{In eq.~\eqref{eq:S_WZ2} we implicitly project the left hand sides on their components of rank $p+1$.}
\begin{eqaed}\label{eq:C-STrF}
    C \wedge \STr \left(i \, \mathcal F \right) & = C_{(p-1)} \wedge \left( F^{(1)} - F^{(2)} \right) \, ,   \\
    C \wedge \STr \left(i \, \mathcal F \wedge i \, \mathcal F \right) & = C_{(p-3)} \wedge \left( F^{(1)} \wedge F^{(1)} - F^{(2)} \wedge F^{(2)} \right)\\
& \quad{} + C_{(p-1)} \wedge \big( -2 \, \beta^2 \, \abs{T}^2 \left( F^{(1)} - F^{(2)} \right) \\ & \quad{} 
    + 2 \, i \, \beta^2 \, DT \wedge (DT)^* \big) \, , \\
    C \wedge \STr \left(i \, \mathcal F \wedge i \, \mathcal F \wedge i \, \mathcal F \right) & = C_{(p-5)} \wedge \left( F^{(1)} \wedge F^{(1)} \wedge F^{(1)} - F^{(2)} \wedge F^{(2)} \wedge F^{(2)} \right) \\
    & \quad{} + C_{(p-3)} \wedge \left( -3 \, \beta^2 \, \abs{T}^2 \big( F^{(1)} \wedge F^{(1)} - F^{(2)} \wedge F^{(2)} \right)   \\
    &\quad{} + 3i \, \beta^2 \left( F^{(1)} + F^{(2)} \right) \wedge DT \wedge (DT)^* \big) \\
    & \quad{}+ C_{(p-1)} \wedge \big( 3 \, \beta^4 \, \abs{T}^4 \wedge \left( F^{(1)} - F^{(2)} \right) \\
    & \quad{}
    - 6 \, i \, \beta^4 \, \abs{T}^2 \, DT \wedge (DT)^* \big) \, .
\end{eqaed}

The inclusion of tachyons in the WZ action was also discussed in~\cite{Kraus:2000nj,Takayanagi:2000rz}. We remark that a more complete description would be especially relevant for non-BPS brane combinations, since the preceding expressions arise from the low-energy expansion of corresponding scattering amplitudes. Setting the tachyon field to zero reduces the resulting action to the standard one pertaining to BPS brane systems.

\section{Computation of the Five-Point Amplitude}\label{sec:higher}

In this section we examine the S-Matrix element for a closed-string RR state, a world-volume scalar, a world-volume gauge boson and two tachyons. This arises from the correlator 
\begin{equation}\label{eq:vertex_correlator_CFT}
    \langle V_{C^{-1}}(z,\bar z)V_{\phi^{0}} (x_1)V_{A^{-1}}(x_2)V_{T^{0}}(x_3)V_{T^{0}}(x_4)\rangle \, .
\end{equation}
We proceed to expand the resulting S-Matrix element in a specific limit, investigating the structure of its vector, (massless) scalar and tachyonic singularities, and we match the singularity structure to the corresponding EFT interactions. We shall focus on bulk singularity structures, where resonances carry RR momentum in the transverse directions.

\subsection{Setup}\label{sec:setup}
The following correlation functions enter the computation of the amplitude:
\begin{eqaed}\label{eq:I1_I2}
    I_1^{cba} & = \langle : S_\alpha(z) : S_\beta(\bar{z}) : \psi^a(x_2) : \psi^b(x_3) : \psi^c(x_4) : \rangle \, , \\
    I_2^{cbaid} & = \langle : S_\alpha(z) : S_\beta(\bar{z}) : \psi^d(x_1) \psi^i(x_1) : \psi^a(x_2) : \psi^b(x_3) : \psi^c(x_4) : \rangle \, .
\end{eqaed}
Specifically, $I_1^{cba}$ is
\begin{eqaed}\label{eq:I1_expression}
I_1^{cba} & = \bigg\{(\Gamma^{cba} \, C^{-1})_{{\alpha\beta}}-\alpha' \eta^{ab}(\gamma^{c} \, C^{-1})_{\alpha\beta} \, \frac{\Re(x_{25}x_{36})}{x_{23}x_{56}} +\alpha' \eta^{ac}(\gamma^{b} \, C^{-1})_{\alpha\beta} \, \frac{\Re(x_{25}x_{46})}{x_{24}x_{56}}\\&\quad{}-\alpha' \eta^{bc}(\gamma^{a} \, C^{-1})_{\alpha\beta} \, \frac{\Re(x_{35}x_{46})}{x_{34}x_{56}}\bigg)\bigg\} \, 2^{-\frac{3}{2}} \, x_{56}^{\frac{1}{4}} \, (x_{25}x_{26}x_{35}x_{36}x_{45}x_{46})^{-\frac{1}{2}} \, ,
\end{eqaed}
where we have introduced the variables $x_5 \equiv z = x + i y \, , x_6 \equiv \bar z$ and $x_{ij} = x_i - x_j$, while $I_2^{cbaid}$ is
\begin{eqaed}\label{eq:I2_expression}
I_2^{cbaid} & =
\bigg\{(\Gamma^{cbaid} \, C^{-1})_{{\alpha\beta}} \\& \quad{} - \frac{1}{2} \, \alpha' \, l_1 \, \frac{\Re(x_{15}x_{26})}{x_{12}x_{56}} - \frac{1}{2} \,
\alpha' \, l_2 \, \frac{\Re(x_{15}x_{36})}{x_{13}x_{56}} + \frac{1}{2} \, \alpha' \, l_3 \, \frac{\Re(x_{15}x_{46})}{x_{14}x_{56}}\\& \quad{} - \frac{1}{2} \, \alpha' \, l_4 \, \frac{\Re(x_{25}x_{36})}{x_{23}x_{56}} + \frac{1}{2} \,
\alpha' \, l_5 \, \frac{\Re(x_{25}x_{46})}{x_{24}x_{56}} - \frac{1}{2} \, \alpha' \, l_6 \, \frac{\Re(x_{35}x_{46})}{x_{34}x_{56}}
\\&\quad{}
- \frac{1}{4} \, (\alpha')^2 \, l_7\bigg(\frac{\Re(x_{15}x_{26})}{x_{12}x_{56}}\bigg)\bigg(\frac{\Re(x_{35}x_{46})}{x_{34}x_{56}}\bigg) \\& \quad{}
+ \frac{1}{4} \, (\alpha')^2 \, l_8
\bigg(\frac{\Re(x_{15}x_{36})}{x_{13}x_{56}}\bigg)\bigg(\frac{\Re(x_{25}x_{46})}{x_{24}x_{56}}\bigg)
\\&\quad{}
- \frac{1}{4} \, (\alpha')^2 \, l_9 \bigg(\frac{\Re(x_{15}x_{46})}{x_{14}x_{56}}\bigg)\bigg(\frac{\Re(x_{25}x_{36})}{x_{23}x_{56}}\bigg)
\bigg\} \\& \quad{}
\times 2^{-\frac{5}{2}} \, x_{56}^{\frac{5}{4}} \, (x_{25}x_{26}x_{35}x_{36}x_{45}x_{46})^{-\frac{1}{2}} \, 
(x_{15}x_{16})^{-1} \, .
\end{eqaed}
For brevity, we have also defined the structures~\footnote{Here we suppress indices for convenience.}
\begin{eqaed}\label{eq:l_structures}
l_1 & = - \, 2 \, \eta^{ad}(\Gamma^{cbi} \, C^{-1})_{\alpha\beta}
\, , & l_2 & = 2 \, \eta^{db}(\Gamma^{cai} \, C^{-1})_{\alpha\beta} \, , & l_3 & = 2 \, \eta^{dc}(\Gamma^{bai} \, C^{-1})_{\alpha\beta} \, , \\
l_4 & = 2 \, \eta^{ab}(\Gamma^{cid} \, C^{-1})_{\alpha\beta} \, , & l_5 & = 2 \, \eta^{ac}(\Gamma^{bid} \, C^{-1})_{\alpha\beta}\,, &
l_6 & = 2 \, \eta^{bc}(\Gamma^{aid} \, C^{-1})_{\alpha\beta} \, , \\
l_7 & = 4 \, \eta^{ad}\eta^{bc}(\gamma^{i} \, C^{-1})_{\alpha\beta} \, , &
l_8 & = 4 \, \eta^{db}\eta^{ac}(\gamma^{i} \, C^{-1})_{\alpha\beta} \, , &
l_9 & = 4 \, \eta^{dc}\eta^{ab}(\gamma^{i} \, C^{-1})_{\alpha\beta} \, .
\end{eqaed}

\subsection{The Final Amplitude}\label{sec:amplitude}

A tedious calculation yields the S-Matrix element of interest, which reads
\begin{eqaed}\label{eq:amplitude1}
\A^{C^{-1} \phi^0 A^{-1} T^0 T^0} & \propto 2 \, \Tr(\lambda_1 \lambda_2 \lambda_3 \lambda_4) \, (P_{-}\fsH_{(n)}M_p)^{\alpha \beta} \, \xi_{1i} \, \xi_{2a} \\
& \quad{}\times\int dx_1 dx_2
dx_3 dx_4 dx_5 dx_6 \, I  \, x_{56}^{-\frac{1}{4}} \, (x_{25}x_{26})^{-\frac{1}{2}} \\
&\quad{}\times\left( i \, (\alpha')^2 \, k_{3b} \, k_{4c} \,  p^i \, \frac{x_{56}}{x_{15}x_{16}} \, I_1^{cba} - i \, (\alpha')^3 \, k_{1d} \, k_{3b} \, k_{4c} \, I_2^{cbaid} \right) \, ,
\end{eqaed}
where
\begin{eqaed}\label{eq:I_factor}
I & = |x_{12}|^{-2t}|x_{13}|^{-2s-\frac{1}{2}}|x_{14}|^{-2v-\frac{1}{2}}|x_{23}|^{-2u-\frac{1}{2}}
|x_{24}|^{-2r-\frac{1}{2}}|x_{34}|^{-2w-1}|x_{15}x_{16}|^{t+s+v+\frac{1}{2}} \\
& \quad{} \times
|x_{25}x_{26}|^{t+u+r+\frac{1}{2}}|x_{35}x_{36}|^{s+u+w+\frac{1}{2}}|x_{45}x_{46}|^{v+r+w+\frac{1}{2}}|x_{56}|^{-2(s+t+u+v+r+w)-2} \, .
\end{eqaed}
We also define the six independent Mandelstam variables,
\begin{eqaed}\label{eq:mandelstam_vars}
& s = - \left(\frac{1}{4} + 2 \, k_1 \cdot k_3 \right) \, , & t &= - 2 \, k_1 \cdot k_2 \, , & v = - \left( \frac{1}{4} + 2 \, k_1 \cdot k_4 \right) \, , \\
& u = - \left( \frac{1}{4} + 2 \, k_2 \cdot k_3 \right) \, , & r & = - \left( \frac{1}{4} + 2 \, k_2 \cdot k_4 \right) \, , & w = - \left( \frac{1}{2} + 2 \, k_3 \cdot  k_4 \right) \, .
\end{eqaed}

The amplitude has been presented in a form that makes $SL(2,\mathbb{R})$ invariance manifest. Hence, one may fix three positions on the world-sheet~\footnote{Closed-string positions are to be counted twice in this respect, and their positions can be fixed anywhere on the upper-half complex plane, including infinity.}, conformally mapped to the upper-half complex plane, taking into account that open string insertions live on the boundary and are thus to be ordered. We make the choice
\begin{equation}\label{eq:gauge_fixing}
    x_1 = 0 \, , \qquad  0 \leq x_2 \leq 1 \, , \qquad  x_3 = 1 \, , \qquad x_4 = \infty \, ,
\end{equation}
which brings the amplitude to the form
\begin{eqaed}\label{eq:amplitude2}
    & \A^{C^{-1} \phi^0 A^{-1} T^0T^0} \propto - \, 4 \, i \, \xi_{1i} \, \xi_{2a} \,  2^{-\frac{1}{2}} \, (P_{-}\fsH_{(n)}M_p)^{\alpha\beta}\int_0^1 dx_2 \, x_2^{-2t} \, (1-x_2)^{-2u-\frac{1}{2}}\int_{\Hp} d^2 z \\ & \quad{} \times |1-z|^{2s+2u+2w} \, |z|^{2t+2s+2v-1} \, k_{3b} \, k_{4c} \, (z - \bar{z})^{-2(t+s+u+v+r+w)-2} \, |x_2-z|^{2t+2u+2r-1} \\ & \quad{}
\times \bigg\{- p^i \bigg((z-\bar z)(\Gamma^{cba} \, C^{-1})_{{\alpha\beta}} + 2 \, \eta^{ab} \, (\gamma^{c} \, C^{-1})_{\alpha\beta} \, \frac{x_2-x \, x_2-x+|z|^{2}}{(1-x_2)} \\ & \quad\quad{}
+ 2 \, \eta^{ac} \, (\gamma^{b} \, C^{-1})_{\alpha\beta}(x-x_2)+2 \, \eta^{bc} \, (\gamma^{a} \, C^{-1})_{\alpha\beta}(1-x)\bigg)\\
&\quad\quad{} + k_{1d} \, \bigg[ (z - \bar{z})(\Gamma^{cbaid} \, C^{-1})_{\alpha\beta}+l_1 \, \frac{-x \, x_2+|z|^{2}}{x_2} + l_2 \left(- x+|z|^{2} \right) + l_3 \, x \\
&\quad\quad{} + l_4 \, \frac{x_2-x \, x_2-x+|z|^{2}}{(1-x_2)} + l_5 \left(x-x_2\right) + l_6 \left(1-x\right) \\ & \quad\quad{}
+ (z - \bar{z})^{-1}\bigg( l_7 \, (-1+x) \, \frac{-x \, x_2+|z|^{2}}{x_2} + l_8 \left(-x+|z|^{2} \right) \left(x_2-x \right) \\ & \quad\quad{}
+ l_9 \, x \, \frac{x_2-x \, x_2-x+|z|^{2}}{(1-x_2)} \bigg)\bigg]\bigg\}  \, \Tr(\lambda_1 \lambda_2 \lambda_3 \lambda_4) \, ,
\end{eqaed}
and the integration over $x_2$ is consistent with the ordering on the real axis.

The total amplitude can be expressed in terms of the (analytic continuation of the) integrals 
\begin{eqaed}\label{eq:total_amplitude_def}
    \A_{xz}(a,b,c,d\mid \alpha,\beta\mid\epsilon) & \equiv \int_\Hp d^2 z \int_0^1 dx\, \abs{1-z}^a \abs{z}^b \,(z-\bar{z})^c (z+\bar{z})^d   \\ 
    & \times (1-x)^\alpha x^\beta\, \abs{x - z}^\epsilon \, ,
\end{eqaed}
whose properties will be described in detail in Section~\ref{sec:beyond}. Letting
\begin{eqaed}\label{eq:F_int}
    \Fxz{\alpha}{\beta}{b}{c}{d} & \equiv 2^{-d} \, \A_{xz}\bigg(2s+2u+2w,2t+2s+2v-1+b, \\ & \quad{}
    -2(t+s+u+v+r+w)-2+c,d \,\bigg| \\ & \quad{}
     -2u-\frac{1}{2}+\alpha,-2t+\beta \,\bigg| \, 2t+2u+2r-1\bigg) \, , 
\end{eqaed}
the total amplitude $\A$ takes the form
\begin{equation}\label{eq:total_amplitude_rewrite}
    \A \propto - \, 4 \, i \, \xi_{1i} \,  \xi_{2a} \,  2^{-\frac{1}{2}} \, (P_{-}\fsH_{(n)}M_p)^{\alpha\beta} \, k_{3b} \, k_{4c} \, \mathcal{Q}^{iabc}_{\alpha \beta}  \, \Tr(\lambda_1 \lambda_2 \lambda_3 \lambda_4) \, ,
\end{equation}
where
\begin{eqaed}\label{Q_term}
    \mathcal{Q}^{iabc}_{\alpha \beta} & \equiv \left(-p^i (\Gamma^{cba}C^{-1})_{\alpha \beta} + k_{1d}(\Gamma^{cbaid}C^{-1})_{\alpha \beta} \right)\Fxz{0}{0}{0}{1}{0} \\
    & \quad{} + \left(-2p^i\eta^{ab}(\gamma^c C^{-1})_{\alpha \beta} + k_{1d}l_4  \right)\left(\Fxz{-1}{1}{0}{0}{0} - \Fxz{-1}{1}{0}{0}{1} - \Fxz{-1}{0}{0}{0}{1} + \Fxz{-1}{0}{2}{0}{0} \right) \\
    & \quad{} + \left(-2p^i \eta^{ac}(\gamma^b C^{-1})_{\alpha \beta} + k_{1d} (l_3-l_2+l_5-l_6-l_1) +2p^i \eta^{bc} (\gamma^a C^{-1})_{\alpha \beta}\right)\Fxz{0}{0}{0}{0}{1} \\
    & \quad{} - \left(-2p^i \eta^{ac} (\gamma^b C^{-1})_{\alpha \beta} + k_{1d}l_5 \right) \Fxz{0}{1}{0}{0}{0} + \left(-2p^i \eta^{bc} (\gamma^a C^{-1})_{\alpha \beta} + k_{1d}l_6 \right) \Fxz{0}{0}{0}{0}{0}\\
    & \quad{} + k_{1d} \Big[ l_1 \Fxz{0}{-1}{2}{0}{0} + l_2 \Fxz{0}{0}{2}{0}{0} + l_7 (\Fxz{0}{0}{0}{-1}{1} - \Fxz{0}{0}{0}{-1}{2} - \Fxz{0}{-1}{2}{-1}{0} + \Fxz{0}{-1}{2}{-1}{1}) + \\
    & \quad{} + l_8 (\Fxz{0}{0}{0}{-1}{2} - \Fxz{0}{1}{0}{-1}{1} + \Fxz{0}{1}{2}{-1}{0} - \Fxz{0}{0}{2}{-1}{1}) \\
    & \quad{} + l_9 (\Fxz{-1}{1}{0}{-1}{1} - \Fxz{-1}{1}{0}{-1}{2} - \Fxz{-1}{0}{0}{-1}{2} + \Fxz{-1}{0}{2}{-1}{1}) \Big] \, .
\end{eqaed}
One can show that this amplitude vanishes~\footnote{In the traces involving $\gamma^{11}$, the special property $H_{(n)} = \star \, H_{(10-n)}$ holds for $n \geq 5$.} unless the rank of the RR potential is $p-3$, $p-1$ or $p+1$.

\subsection{Factorized Limit}\label{sec:factorized_limit}

Let us begin by discussing the limit $4 \, k_2 \cdot p \rightarrow 1$ that was studied in~\cite{Hatefi:2018qlm}. In this case the amplitude simplifies considerably, since the integrand decouples in the $z \, , \bar z$ and $x$ variables and the integrals factorize. The resulting reduced integrals can be written in closed form, and are presented in~\cite{Hatefi:2012wj,Fotopoulos:2001pt}. For completeness, their derivation is reproduced in Appendix~\ref{appendix:z_integral}. One can write
\begin{equation}\label{eq:A1+2+3+4}
    \A = (\A_1 + \A_2 +\A_3 + \A_4)\,\Tr(\lambda_1\lambda_2\lambda_3\lambda_4) \, ,
\end{equation}
where $\A_1$ is only non-vanishing for $C_{p-3}$ and reads
\begin{eqaed}\label{eq:A1}
    {\A}_{1} & \rightarrow \frac{64 \, i \, \xi_{1i} \, \xi_{2a} \, 2^{-\frac{1}{2}} \, \pi \, N_1 \, N_2 \, k_{3b} \, k_{4c}}{(p-2)!}\bigg(k_{1d} \, \epsilon^{a_0\ldots a_{p-4}cbad} H^{i}_{a_0 \ldots a_{p-4}} \\
    & \quad{} - p^i \,  \epsilon^{a_0 \ldots a_{p-3}cba} H_{a_0\ldots a_{p-3}}\bigg) \, .
\end{eqaed}
Moreover, $\A_2 \, , \A_3$ are only non-vanishing for $C_{p-1}$, and read
\begin{eqaed}\label{eq:A2}
 \A_2 & \rightarrow \frac{64 \, i \, \xi_{1i} \, 2^{-\frac{1}{2}} \, \pi \, N_3 \, N_4 }{p!} \, \epsilon^{a_0...a_{p-1}a} \\
 & \quad{} \times
\bigg[ \left( 2 \, k_3 \cdot \xi_2 \, p^i \, k_{4a} \, H_{a_0...a_{p-1}} - 2 \, k_3 \cdot \xi_2 \, k_{4a} \, H^{i}_{a_0...a_{p-2}} \, k_{1a_{p-1}} \right) Q_1 \\
& \quad{} + \left(- 2 \, k_4 \cdot \xi_2 \, p^i \, k_{3a} \, H_{a_0...a_{p-1}} + 2 \, k_4 \cdot \xi_2 \, k_{3a} \, H^{i}_{a_0...a_{p-2}} \, k_{1a_{p-1}}\right) Q_2 \\
& \quad{}
+ \left( p^i \left(w + \frac{1}{2} \right) \xi_{2a} \, H_{a_0...a_{p-1}} - \left(w + \frac{1}{2} \right) H^{i}_{a_0...a_{p-2}} \, k_{1a_{p-1}} \, \xi_{2a} \right) Q_3 \bigg] \, ,
\end{eqaed}
\begin{eqaed}\label{eq:A3}
 \A_3 & \rightarrow \frac{64 \, i \, \xi_{1i} \, 2^{-\frac{1}{2}} \, \pi \, N_5 \, N_6 }{p!} \, \epsilon^{a_0...a_{p-2}ca} H^{i}_{a_0...a_{p-2}}
\bigg[ 2 \, k_1 \cdot \xi_2 \, k_{3a} \, k_{4c} \, Q_4 \\
& \quad{} - \left( s + \frac{1}{4} \right) k_{4c} \, \xi_{2a} \, Q_5 - \left( v + \frac{1}{4} \right) k_{3c} \, \xi_{2a} \, Q_6 \bigg] \, .
\end{eqaed}
Finally, $\A_4$ is non-vanishing only for $C_{p+1}$, and reads
\begin{equation}\label{eq:A4}
  \A_4 \rightarrow \frac{64 \, i \, \xi_{1i} \, 2^{-\frac{1}{2}} \, \pi \, N_7 \, N_8 }{(p+1)!} \, \epsilon^{a_0...a_{p}} H^{i}_{a_0...a_{p}}
\bigg[ 2 \, k_1 \cdot \xi_2 \, Q_7 + 2 \, k_4 \cdot \xi_2 \, Q_8 + 2 \, k_3 \cdot \xi_2 \, Q_9 \bigg] \, .
\end{equation} 
The functions $N_i,Q_i$, used to simplify the notation, are listed in Appendix~\ref{appendix:qandns}.

\subsection{Vector Bulk Singularity Structure}\label{sec:vector_bulk_sing}

In order to investigate the structure of poles in the amplitude, one must consider the correct limit for the Mandelstam variables around which to set up a low-energy expansion. It turns out that the proper regime is
\begin{eqaed}\label{eq:limittsvpw}
t \rightarrow  0 \, , \quad s \, , v \, , - \, p^a p_a  \rightarrow  - \frac{1}{4} \,,\quad  w \rightarrow  - \frac{1}{2} \, ,\\
\frac{1}{2}\left[(u\rightarrow 0 \, , r \rightarrow -\frac{1}{4}) + (u\rightarrow -\frac{1}{4} \, , r
\rightarrow 0) \right] \, .
\end{eqaed}
The notation in the last line of eq.~\eqref{eq:limittsvpw} prescribes that the S-Matrix be averaged over the two limits. Eq.~\eqref{eq:limittsvpw} is consistent with momentum conservation,
\begin{equation}\label{eq:mandelstam_sum}
    s+t+u+v+r+w = - \, p^a p_a -1 \, ,
\end{equation}
and conforms to the symmetries of the S-Matrix, namely invariance under the interchanges of $u$ with $r$ and of $s$ with $v$. The limit $-p^a p_a \rightarrow -\frac{1}{4}$ specifically, instead, conforms to the RR-tachyon two-point function, for which $-p^a p_a = -\frac{1}{4}$ is a constraint.

As an example, $\A_3$ has massless $t$-channel poles, and in the specified limit it expands as~\cite{Hatefi:2018qlm}
\begin{equation}\label{A3_pole}
\A_3 = - \, \frac{\pi^{3/2}}{t} - \, \frac{\pi^{7/2}}{6t} \left[ (s+u+v+r)^2 + \ldots \right] + \ldots \, ,
\end{equation}
where the contributions inside the brackets involve only Mandelstam variables different from $t$, and correct the residue with higher-derivative terms, while the ellipses outside the brackets indicate non-singular terms.

\begin{figure}
\begin{center}
\begin{tikzpicture}
  \begin{feynman}[small]
    \vertex (a);
    \vertex [right=2cm of a] (b);
    \vertex [above left=of a] (i1) {$T_3$};
    \vertex [below left=of a] (i2) {$T_4$};
    \vertex [above right=of b] (o1) {$C_{(p-3)}$};
    \vertex [right=of b] (o2) {$A_2$};
    \vertex [below right=of b] (o3) {$\phi_1$};
    \diagram* {
    (i1) -- [plain] (a) -- [plain] (i2),
    (a)-- [photon, edge label=$A$, momentum'=$w$] (b),
    (o1) -- [gluon] (b) -- [photon] (o2),
    (o3) -- [plain] (b)
    };
  \end{feynman}
\end{tikzpicture}
\end{center}
\caption{A Feynman diagram representation of the massless vector $w$-channel pole of eq.~\eqref{eq:eft_subamplitude}.}
\label{fig:feynman_A}
\end{figure}
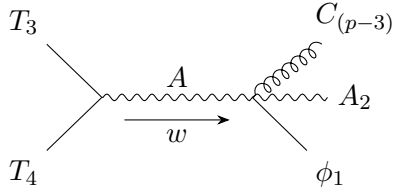

We now describe in detail how to extract the singularity structure of the S-Matrix, and in particular the contributions of resonances carrying RR momentum in the transverse directions, which we denote collectively ``bulk singularity structure''. To begin with, let us consider poles associated to the gauge field. In the $n=p-2$ case the first gauge field pole in $\A_1$ is
\begin{equation}\label{eq:gaugepole}
    \frac{64 \, i \, \pi^{3/2}}{(p-2)!} \, \frac{1}{w} \, \xi_{1i} \, \xi_{2a} \, k_{3b} \, k_{4c} \, k_{1d} \, \epsilon^{a_0\ldots a_{p-4}cbad} H^{i}_{a_0 \ldots a_{p-4}}  \, \Tr(\lambda_1\lambda_2\lambda_3\lambda_4)
\end{equation}
where the $\lambda_i$ are CP factors.
This identifies, on the EFT side, an interaction term of the form
\begin{equation}\label{eq:eft_subamplitude}
     V^{\alpha}_{a}(T_3,T_4,A) \,  G^{\alpha\beta}_{ab}(A) \, \tilde{V}^{\beta}_{b}(C_{p-3},A,\phi_1,A_2) \, ,
\end{equation}
determined by the Feynman rules
\begin{eqaed}\label{eq:feynman_rules1}
V^{\alpha}_{a}(T_3,T_4,A) & = i \, T_p \, (2\pi\alpha') \, (k_3-k_4)_{a} \, \Tr(\lambda_3\lambda_4\lambda^{\alpha})\\
G^{\alpha\beta}_{ab}(A) & = \frac{i \, \delta^{ab} \, \delta^{\alpha\beta}}{(2\pi\alpha')^2 \, T_p} \, \frac{1}{w} \\
{V}^{\beta}_{b}(C_{p-3},A,\phi_1,A_2)&{}= i \, \mu'_p \, \beta \, \frac{(2\pi\alpha')^3}{(p-2)!} \,\epsilon^{a_0\cdots a_{p-1}b} \\
& \quad{} \times H^{i}_{a_0\cdots a_{p-4}} \, k_{1a_{p-3}} \, k_{2a_{p-2}} \, \xi_{2a_{p-1}} \, \xi_{1i} \, \Tr(\lambda_1\lambda_2\lambda^{\beta})
\end{eqaed}
for the vertices and propagators, where $\alpha,\beta$ are indices in the adjoint representation of the gauge group. The sub-amplitude is depicted in Fig.~\ref{fig:feynman_A}.

In order to reconstruct the pole in eq.~\eqref{eq:gaugepole} from this term, we enforce momentum conservation in directions parallel to the branes, $(k_1 + k_2 + k_3 + k_4 + p)^a=0$, in order to eliminate $k_4$ in $V_a^\alpha$. We then employ the Bianchi identity
\begin{equation}\label{eq:bianchi2}
      p_{b} \, H_{a_0\cdots a_{p-4}} \epsilon^{a_0\cdots a_{p-1}b}=0\,,
\end{equation}
and take into account that $k_{2b}k_{2a_{p-2}} \epsilon^{a_0\ldots a_{p-1}b}$ and $k_{1b}k_{1a_{p-2}} \epsilon^{a_0\ldots a_{p-1}b}$ do not contribute to the pole, since the relevant S-Matrix residue is symmetric under the exchange of $k_1$ and $k_2$. Up to a normalization factor of $\frac{1}{2}\mu'_p\beta \sqrt{\pi}$, one finds in this fashion that the S-Matrix pole is reconstructed from the EFT contribution.

Let us also observe that the vertex $\tilde{V}^{\beta}_{b}(C_{p-3},A,\phi_1,A_2)$ arises from the so-called mixed Chern-Simons and Taylor expansion coupling
\begin{equation}\label{eq:chernsimonstaylor}
     \beta \, \mu'_p \, (2\pi\alpha')^3 \, \int_{\Sigma_{(p+1)}} \!\!\!\!\!\!\! d^{p+1} \sigma \, \Tr \left( \partial_i C_{p-3}\wedge F\wedge F\, \phi^i \right) \, .
\end{equation}
Let us now focus specifically on the bulk singularity structure, that is on poles that carry transverse RR momentum. In the present case, the relevant terms are those containing $p\cdot \xi_1$, and the first bulk pole in the S-Matrix takes the form
\begin{equation}\label{eq:bulkpole}
     \frac{32 \, i \, \beta \, \mu'_p \, \pi^2 \, \xi_{2a} \, k_{3b} \, k_{4c}}{(p-2)!} \, \frac{1}{w} \, (p \cdot \xi_1) \, \epsilon^{a_0\ldots a_{p-3}cba} H_{a_0\ldots a_{p-3}} \, .
\end{equation}
We now show how it is recovered from an EFT coupling. Integrating eq.~\eqref{eq:chernsimonstaylor} by parts and assuming fast decay at infinity~\footnote{While at the level of six-point correlators only partial derivatives appear, gauge invariance dictates that higher-point contributions will reconstruct minimal couplings.} one finds
\begin{eqaed}\label{eq:chernsimonstaylor_byparts}
   -\beta \, \mu'_p \, (2\pi\alpha')^3 \int_{\Sigma_{(p+1)}}\!\!\!\!\!\!\! d^{p+1} \sigma \, \epsilon^{a_0...a_p} & \Tr\bigg( \partial_{a_{p-3}} \partial_i C_{a_0...a_{p-4}} \, A_{a_{p-2}} \, D_{a_{p-1}} A_{a_{p}} \, \phi^i \\
   & - \partial_i C_{a_0...a_{p-4}} \, A_{a_{p-2}} \, D_{a_{p-1}} 
  A_{a_{p}} \, D_{a_{p-3}} \phi^i \bigg) \, ,
\end{eqaed}
and the second term in eq.~\eqref{eq:chernsimonstaylor_byparts} yields the aforementioned pole of eq.~\eqref{eq:gaugepole}. Upon the substitution
\begin{equation}\label{eq:dC-H}
    (p-3) \, \partial_i C_{a_0\ldots a_{p-4}}= H_{i\,a_0 \ldots a_{p-4}}-\partial_{[a_{p-4}} C_{a_0\ldots a_{p-5}]i} \, ,
\end{equation}
the first term indeed generates the bulk singularity~\eqref{eq:bulkpole}.

There are also additional higher-order bulk singularity structures to be identified, and we delineate the following strategy for determining all these contributions. To begin with, note that the kinetic term $(2\pi\alpha') D^aT D_aT$ of the tachyon does not receive corrections, since it is fixed by canonical normalization, and therefore neither does the $V^\alpha_a(T_3,T_4,A)$ vertex arising from minimal coupling. Moreover, the gauge kinetic term is also fixed, so that $G^{\alpha\beta}_{ab}(A)$ receives no corrections. Hence, in order to determine the rest of the bulk poles in the particular amplitude, it suffices to look at the vertex $V^{\beta}_b(C_{p-3},A,\phi_1,A_2)$ and, correspondingly, at the series of higher-order corrections to~\eqref{eq:chernsimonstaylor}, which takes the form
\begin{equation}\label{eq:chernsimonstaylor_allorders}
    \beta \, \mu'_p \, (2\pi\alpha')^3 \sum_{n=-1}^{\infty}a_n
 \int_{\Sigma_{(p+1)}} \!\!\!\!\!\!\! d^{p+1} \sigma \, \Tr \left(\partial_i C_{p-3}\wedge D^{a_1}\cdots D^{a_{n+1}}F\wedge D_{a_1}\cdots D_{a_{n+1}}(F \, \phi^i)\right) \, .
\end{equation}
The all-order modified vertex $\tilde{V}^{\beta}_{b}(C_{p-3},A,\phi_1,A_2)$ that can be extracted from eq.~\eqref{eq:chernsimonstaylor_allorders} would then allow, after substitution into the sub-amplitude in eq.~\eqref{eq:eft_subamplitude}, to reconstruct the entire bulk singularity structure.

\subsection{Scalar Bulk Singularity Structure}\label{sec:scalar_bulk_sing}

%Having presented the technique for extracting $t$-channel poles, we now move onto examining the singularity structure in the other channels.

Let us now move on to the singularity structure of scalar poles (massless and tachyonic). To begin with, we observe that selection rules forbid an $A \, \phi \, T \, T$ coupling. Specifically, the relevant operator product includes the trace of a product of Pauli matrices that vanishes in this case, thereby excluding the presence of the corresponding coupling. This is well-reflected in the EFT, where the amplitude vanishes on-shell by Lorentz invariance~\footnote{Here we refer to the residual Lorentz symmetry in the presence of D-branes.}. The preceding argument shows that this coupling is still absent even after the inclusion of higher-derivative corrections. Consequently, there are no poles in a $(t+v'+r')$-channel, with $v'=v+\frac{1}{4}$, $r'=r+\frac{1}{4}$, as discussed in~\cite{Hatefi:2016yhb}.

%\RA{What is the point of the above comment? Could it be moved down to after the presentation of the DBI action?}

It is interesting to examine the effective corrections to the DBI sector of the action. The relevant portion of the DBI action is~\cite{Hatefi:2016yhb}
\begin{eqaed}\label{eq:DBI_expanded}
{\cal
L}_{DBI}  & = - \, T_p \, (2\pi\alpha')\left(m^2|T|^2+DT\cdot(DT)^{*}-\frac{\pi\alpha'}{2}
\left(F^{(1)}\cdot{F^{(1)}}+
F^{(2)}\cdot{F^{(2)}}\right)\right)\\
&\quad{}+ T_p \, (\pi\alpha')^3\left(\frac{2}{3}DT\cdot(DT)^{*}\left(F^{(1)}\cdot{F^{(1)}}+F^{(1)}\cdot{F^{(2)}}+F^{(2)}\cdot{F^{(2)}}\right)\right.\\
&\quad{}\left.+\frac{2m^2}{3} \, |T|^2\left(F^{(1)}\cdot{F^{(1)}}+F^{(1)}\cdot{F^{(2)}}+F^{(2)}\cdot{F^{(2)}}\right)\right.\\
&\quad{}\left.\frac{4}{3}\left((D^{\mu}T)^*D_{\beta}T+D^{\mu}T(D_{\beta}T)^*\right)\left({F^{(1)}}^{\mu\alpha}F^{(1)}_{\alpha\beta}+{F^{(1)}}^{\mu\alpha}F^{(2)}_{\alpha\beta}+{F^{(2)}}^{\mu\alpha}F^{(2)}_{\alpha\beta}\right)\right)\,,
\end{eqaed}
where traces over the gauge degrees of freedom are implicit. The $A \, \phi \, \phi$ vertex receives no corrections, since it is fixed by the kinetic term $\frac{1}{2} (2\pi\alpha')^2 D\phi^i D\phi_i $, nor does the $\phi \, \phi$ propagator. Therefore, in order to determine all scalar poles in the $t$-channel it suffices to consider the higher-derivative corrections to the Chern-Simons-Taylor interaction. These singularities are captured from the EFT sub-amplitude
\begin{equation}\label{eq:subamplitude_scalar}
V^{\alpha}_{i}(\phi_1,A_2,\phi) \, G^{\alpha\beta}_{ij}(\phi) \, V^{\beta}_{j}(C_{p-1},\phi,T_3,T_4) \, ,
\end{equation}
determined by the Feynman rules
\begin{eqaed}\label{eq:feynman_rules2}
V^{\alpha}_{i}(\phi_1,A_2,\phi) & = - \, 2 \, i \, T_p \, \xi_{1i} \, (2\pi\alpha')^2 \, (k_1 \cdot \xi_2) \, \Tr(\lambda_1\lambda_2\lambda^{\beta}) \\
G^{ij}(\phi) & = \frac{i \, \delta^{ij} \,  \delta^{\alpha\beta}}{(2\pi\alpha')^2 \, T_p } \, \frac{1}{t} \\
\tilde{V}^{\beta}_{j}(C_{p-1},\phi,T_3,T_4)  & = \mu'_p \, \beta \, \frac{(2\pi\alpha')^3}{p!} \, \epsilon^{a_0\ldots a_{p}}H^{j}_{a_0\ldots a_{p-2}} \, k_{3a_{p}} \, k_{a_{p-1}} \, \Tr(\lambda_3\lambda_4\lambda^{\beta}) \, ,
\end{eqaed}
\begin{figure}
\begin{center}
\begin{tikzpicture}
  \begin{feynman}[small]
    \vertex (a);
    \vertex [right=2cm of a] (b);
    \vertex [above left=of a] (i1) {$\phi_1$};
    \vertex [below left=of a] (i2) {$A_2$};
    \vertex [above right=of b] (o1) {$C_{(p-1)}$};
    \vertex [right=of b] (o2) {$T_4$};
    \vertex [below right=of b] (o3) {$T_3$};
    \diagram* {
    (i1) -- [plain] (a) -- [photon] (i2),
    (a)-- [plain, edge label=$\phi$, momentum'=$t$] (b),
    (o1) -- [gluon] (b) -- [plain] (o2),
    (o3) -- [plain] (b)
    };
  \end{feynman}
\end{tikzpicture}
\end{center}
\caption{A Feynman diagram representation of the massless scalar $t$-channel pole of eq.~\eqref{eq:subamplitude_scalar}.}
\label{fig:feynman_phi}
\end{figure}
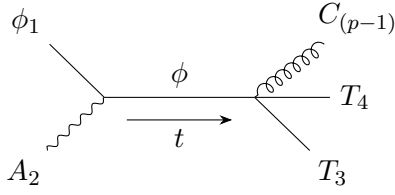
where $k$ is the momentum of the virtual scalar. This sub-amplitude is depicted in Fig.~\ref{fig:feynman_phi}. Repeating the steps followed in the preceding section, starting from the above Feynman rules and using momentum conservation and the Bianchi identity, one can reproduce the scalar $t$-channel pole
\begin{equation}\label{eq:t-pole}
    \frac{64 \, i \, \pi^2}{p!} \, \frac{1}{t} \, \beta \, \mu'_p \, \xi_{1i} \,  (k_1\cdot \xi_2) \, H^{i}_{a_0\ldots a_{p-2}} \, k_{3a_{p}} \, k_{4a_{p-1}} \epsilon^{a_0\ldots a_{p}}\, \Tr(\lambda_1\lambda_2\lambda_3\lambda_4) \, .
\end{equation}
As in the case involving vector poles discussed in the preceding section, the vertex ${V}^{\beta}_{j}(C_{p-1},\phi,T_3,T_4)$ arises from an effective coupling of the type
\begin{equation}\label{eq:cttphidercoupling}
    \beta \, \mu'_p \, (2\pi\alpha')^3 \int_{\Sigma_{p+1}} \!\!\!\!\!\!\! d^{p+1} \sigma \, \Tr \left( \partial_i C_{p-1}\wedge DT\wedge DT \, \phi^i \right) \, .
\end{equation}
Once again, one can recover the entire structure of scalar poles considering the all-order correction to the coupling of eq.~\eqref{eq:cttphidercoupling}, which yields
\begin{eqaed}\label{eq:cttphidercoupling_allorders}
\beta \, \mu'_p \, (2\pi\alpha')^3 \sum_{n=-1}^{\infty} c_n
\int_{\Sigma_{(p+1)}} \!\!\!\!\!\!\! d^{p+1} \sigma \, & \Tr\big[\partial_i C_{p-1}\wedge D^{a_1}\cdots D^{a_{n+1}} DT \\ & \quad
\wedge D_{a_1}\cdots D_{a_{n+1}}(DT \, \phi^i)\big] \, .
\end{eqaed}
Extracting the all-order $\tilde{V}^{\beta}_{j}(C_{p-1},\phi,T_3,T_4)$ vertex from eq.~\eqref{eq:cttphidercoupling_allorders}, and replacing it into the the sub-amplitude eq.~\eqref{eq:subamplitude_scalar}, one can reconstruct the complete series of $t$-channel scalar poles, which reads
\begin{eqaed}\label{eq:all_t-poles}
    \frac{64 \, i \pi^2}{p!} \, \frac{1}{t} \, \beta \, \mu'_p \sum_{n=-1}^{\infty}c_n \, \left(\alpha' \, k \cdot (k_3+k_4)\right)^{n+1} \xi_{1i} \left(k_1.\xi_2\right) & H^{i}_{a_0\ldots a_{p-2}} \, k_{3a_{p}} \, k_{4a_{p-1}} \epsilon^{a_0\ldots a_{p}} \\ & \quad{} \times \Tr(\lambda_1\lambda_2\lambda_3\lambda_4) \, .
\end{eqaed}
This expression can be simplified, since momentum conservation implies
\begin{equation}\label{eq:momentum_conservation_power}
    \left( \alpha' \, k \cdot (k_3+k_4)\right)^{n+1}= (s+v+u+r+1)^{n+1}\,.
\end{equation}
Finally, we can determine the all-order tachyonic singularities in the $u' = u + \frac{1}{4}$ and $r'$ channels, including bulk singularities. After integrating eq.~\eqref{eq:cttphidercoupling_allorders} by parts, which gives
\begin{eqaed}\label{eq:cttphidercoupling_byparts}
     -\beta \, \mu'_p \, (2\pi\alpha')^3 \int_{\Sigma_{(p+1})} \!\!\!\!\!\!\! d^{p+1} \sigma \, \epsilon^{a_0\ldots a_p} & \big( \partial_{a_{p-1}} \partial_i C_{a_0\ldots a_{p-2}} \, T \, D_{a_{p}}T \, \phi^i \\ &
     - \partial_i C_{a_0\ldots a_{p-2}} \, T \, D_{a_{p}} T \, 
 D_{a_{p-1}} \phi^i \big) \, ,
\end{eqaed}
and following the same procedure, we consider the all-order series of corrections to the first term in eq.~\eqref{eq:cttphidercoupling_byparts},
\begin{eqaed}\label{eq:all-order_dC-DT-DTphi}
    - \, \beta \, \mu'_p \, (2\pi\alpha')^3 & \int_{\Sigma_{p+1}}\!\!\!\!\!\!\! d^{p+1} \sigma \, \epsilon^{a_0...a_p} \bigg[ \partial_{a_{p-1}} \partial_i C_{a_0...a_{p-2}}
    \\ & \quad{}
    \sum_{n=-1}^{\infty} D^{d_1}\cdots D^{d_{n+1}} T \,  D_{d_1}\cdots D_{d_{n+1}}\left(D_{a_{p}}T \, \phi^i\right) \bigg]\,,
\end{eqaed}
which again reconstructs the all-order $\tilde{V}^{\beta}_{j}(C_{p-1},T,\phi_1,T_4)$ vertex, to be inserted into the rule
\begin{equation}\label{eq:subamplitude_C-T-phi-T}
V^{\alpha}(T_3,A_2,T) \, G^{\alpha\beta}(T) \, V^{\beta}(C_{p-1},T,\phi_1,T_4) )
\end{equation}
in compliance the Feynman rules
\begin{eqaed}\label{eq:feynman_rules3}
V^{\alpha}_{a}(T_3,A_2,T) & = i \, T_p \, (2\pi\alpha') \, (k_3-k) \cdot \xi_2 \, \Tr(\lambda_2\lambda_3\lambda^{\alpha}) \\
G^{\alpha\beta}(T) & = \frac{i \, \delta^{\alpha\beta}}{(2\pi\alpha') T_p} \, \frac{1}{u'} \\
V^{\beta}_{j}(C_{p-1},T,\phi_1,T_4) & = i \, \mu'_p \, \beta \, \frac{(2\pi\alpha')^3}{p!} \left(p\cdot \xi_1\right) \sum_{n=-1}^{\infty}d_n \, \left(\alpha' \, k \cdot (k_1+k_4)\right)^{n+1} \\
& \quad{} \times H_{a_0...a_{p-1}} \, k_{4a_{p}} \epsilon^{a_0...a_{p}} \, \Tr(\lambda_1\lambda_4\lambda_{\beta}) \, .
\end{eqaed}
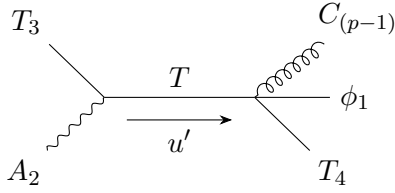
\begin{figure}
\begin{center}
\begin{tikzpicture}
  \begin{feynman}[small]
    \vertex (a);
    \vertex [right=2cm of a] (b);
    \vertex [above left=of a] (i1) {$T_3$};
    \vertex [below left=of a] (i2) {$A_2$};
    \vertex [above right=of b] (o1) {$C_{(p-1)}$};
    \vertex [right=of b] (o2) {$\phi_1$};
    \vertex [below right=of b] (o3) {$T_4$};
    \diagram* {
    (i1) -- [plain] (a) -- [photon] (i2),
    (a)-- [plain, edge label=$T$, momentum'={$u'$}] (b),
    (o1) -- [gluon] (b) -- [plain] (o2),
    (o3) -- [plain] (b)
    };
  \end{feynman}
\end{tikzpicture}
\end{center}
\caption{A Feynman diagram representation of the tachyonic scalar $u'$-channel or $r'$-channel pole of eq.~\eqref{eq:subamplitude_C-T-phi-T}.}
\label{fig:feynman_tachyon}
\end{figure}
This sub-amplitude is represented pictorially in the diagram of Fig.~\ref{fig:feynman_tachyon}

Because of the $p\cdot \xi_1$ coefficient, this contributes to the bulk singularity structure. Indeed, from the above expressions, one can extract all $u'$ and $r'$ tachyonic singularity structures, including bulk poles.

To conclude, we have presented and tested a generic methodology for the reconstruction of all-order EFT couplings for the $\text{D}p$-$\overline{\text{D}p}$ system from superstring (perturbative) scattering amplitudes. We remark, however, that we performed the analysis expanding near the $- \, p^a p_a \rightarrow - \, \frac{1}{4}$ point, as suggested by the RR-tachyon two-point function. Therefore the effective couplings determined in this section are only valid when the low-energy expansion is carried out in this fashion, and they cannot be directly compared, for instance, to BSFT results.

\section{Beyond the Factorized Limit}\label{sec:beyond}

Let us now discuss how one can go beyond the factorized limit. To this end, let us recall that the final amplitude involves (analytic continuations of) integrals of the form
\begin{eqaed}\label{eq:total_amplitude}
    \A_{xz}(a,b,c,d\mid \alpha,\beta\mid\epsilon) & = \int_\Hp d^2 z \int_0^1 dx\, \abs{1-z}^a \abs{z}^b \,(z-\bar{z})^c (z+\bar{z})^d   \\ 
    & \quad{} \times (1-x)^\alpha x^\beta\, \abs{x - z}^\epsilon \, ,
\end{eqaed}
where the integration measure is $d^2 z = 2i \, d \Re(z) \, d \Im(z)$ and $\Hp$ is the upper-half complex plane. In this section we work within the convergence domain of the integral, in such a way as to allow an analytic continuation at the end. The factorized limit amounts to setting $\epsilon = 0$ in the mixing term, and leads to the two simpler integrals
\begin{equation}\label{eq:x_amplitude}
    \A_x(\alpha,\beta) \equiv \int_0^1 dx (1-x)^\alpha x^\beta = \Beta(\alpha+1,\beta+1) \, ,
\end{equation}
\begin{equation}\label{eq:z_amplitude}
    \A_z(a,b,c,d) \equiv \int_\Hp d^2 z \, \abs{1-z}^a \abs{z}^b \,(z-\bar{z})^c (z+\bar{z})^d \, ,
\end{equation}
where eq.~\eqref{eq:x_amplitude} is Euler's $\Beta$ function. The latter integral is evaluated in closed form for any $d \in \mathbb{N}$ in eq.~\eqref{eq:z_result}.

In order to address the case $\epsilon \neq 0$, we split the integration over $z$ into two regions: the (upper-half) unit disk $D$ and its complement $\overline{D}$. This defines the integrals $\A_{xz}^D \, , \, \A_{xz}^{\overline D}$, and the factorized counterparts $\A_z^D \, , \A_z^{\overline D}$, which are discussed in Appendix~\ref{appendix:z_integral}. One can then build suitable expansions for each term, inserting a fictitious parameter $\lambda$ in the integrand. Let us now discuss the two cases separately.
%
%\begin{equation}\label{eq:integral_z_split}
%\A_{xz} = \int_{z \in \Hp} \ldots = \int_{\abs{z}<1, z-\bar{z}>0} \ldots \, + \,     \int_{\abs{z}>1, z-\bar{z}>0} \ldots \equiv \A_{xz}^D + \A_{xz}^{\overline D}\,,
%\end{equation}
%
\subsection{Expansion of the Inner Integral}

In order to evaluate $\A_{xz}^D$, we modify the mixing term via the insertion of a fictitious expansion parameter $\lambda$, according to
\begin{equation}\label{eq:lambda_insertion_disk}
    \abs{x-z}^{\epsilon} \quad \longrightarrow \quad \left(k-\lambda(k-\abs{x-z}^2) \right)^{\epsilon/2} 
\end{equation}
where $k$ is a real constant greater than $4$. For $\lambda = 1$ one recovers the original integral, while for $\lambda = 0$ the integral becomes factorized with a $k^{\frac{\epsilon}{2}}$ prefactor. Note that the modified integral $\A_{xz}^D(\lambda)$ is absolutely convergent whenever $\A_{xz}^D(1)$ is, since for $\abs{\lambda} < 1$ the only effect of $\lambda$ is to resolve the (integrable) singularity at $z \sim x$. The modified mixing term in eq.~\eqref{eq:lambda_insertion_disk} is holomorphic in $\lambda$ within the unit disk, and its Taylor series about $\lambda=0$,
%
%\begin{equation}\label{eq:integrand_expansion_inner}
%       \sum_{p,q,r,s=0}^\infty \binom{\epsilon/2}{p+q+r+s} \, \frac{(p+q+r+s)!}{p! \, q! \, r! \, s!}  \, (-1)^{p+s} \, \frac{x^{2q+s} \, \abs{z}^{2r} \, (z+\bar{z})^{s}}{k^{q+r+s-\frac{\epsilon}{2}}} \, \lambda^{p+q+r+s}
%\end{equation}
%
\begin{equation}\label{eq:integrand_expansion_inner}
       \sum_{p,q,r,s=0}^\infty \binom{\epsilon/2}{p,q,r,s,\epsilon/2-p-q-r-s}  \, (-1)^{p+s} \, \frac{x^{2q+s} \, \abs{z}^{2r} \, (z+\bar{z})^{s}}{k^{q+r+s-\frac{\epsilon}{2}}} \, \lambda^{p+q+r+s} \, ,
\end{equation}
has a radius of convergence of $\frac{k}{k-\abs{x-z}^2} > 1$ since $\abs{z}<1$. Therefore, the modified integral $\A_{xz}^D(\lambda)$ is also holomorphic~\footnote{This follows from Morera's theorem, since a loop integral in $\lambda$ inside the unit disk can be interchanged with the integrals over $x$ and $z$ due to absolute convergence.} within the unit disk, and thus can be Taylor expanded in $\lambda$ in this region integrating term by term and using eq.~\eqref{eq:z_integral_disk_solution} to compute the factorized integrals on the (upper-half) disk. One is thus led to
%
%\begin{equation}\label{eq:amplitude_expansion_inner}
%    \A_{xz}^{D} = \sum_{p , q , r , s = 0}^\infty \frac{(-1)^{p+s} \, \Gamma(1+\frac{\epsilon}{2})}{\Gamma(1+\frac{\epsilon}{2}-M)} \, \frac{\A_x(\alpha,\beta+2q+s) \, \A_z^{D}(a,b+2r,c,d+s)}{k^{q+r+s-\frac{\epsilon}{2}} \, p! \, q! \, r! \, s!} \, \lambda^{M} \, ,
%\end{equation}
%
\begin{eqaed}\label{eq:amplitude_expansion_inner}
    \A_{xz}^{D} & = \sum_{p , q , r , s = 0}^\infty (-1)^{p+s} \, \binom{\epsilon/2}{p,q,r,s,\epsilon/2-p-q-r-s} \\
    & \quad{} \times \frac{\A_x(\alpha,\beta+2q+s) \, \A_z^{D}(a,b+2r,c,d+s)}{k^{q+r+s-\frac{\epsilon}{2}}} \, \lambda^{p+q+r+s} \, ,
\end{eqaed}
and one can verify that each of the factorized integrals is absolutely convergent whenever $\A_{xz}^D(\lambda)$ is, and thus whenever $\A_{xz}^D(1)$ is.
As a final comment, let us mention that the arbitrary constant $k>4$ can be in principle tuned for optimal numerical accuracy.

\subsection{Expansion of the Outer Integral}

Let us now discuss the expansion of $\A_{xz}^{\overline D}$. Similarly to the preceding section, we insert the parameter $\lambda$ according to 
\begin{equation}\label{eq:lambda_insertion_outer}
    \abs{x - z}^\epsilon \quad \longrightarrow \quad \left( ( x \, \lambda - z )( x \, \lambda - \bar{z}) \right)^{\epsilon/2} = \abs{z}^\epsilon \left( 1 + \frac{x^2}{\abs{z}^2} \, \lambda^2 - \frac{x(z+\bar{z})}{\abs{z}^2} \, \lambda \right)^{\epsilon/2} \, ,
\end{equation}
so that the integrand, as a function of $\lambda$, is again holomorphic within the unit disk, since the branch points are located outside for all $z,x$ in the domain of integration. Hence, the expansion of the mixing term
%
%\begin{equation}\label{eq:integrand_expansion_outer}
%    \sum_{p , q=0}^\infty \binom{\epsilon/2}{p+q} \binom{p+q}{p} (-1)^q \, \frac{x^{2p+q} \, (z+\bar{z})^q}{\abs{z}^{2p+2q-\epsilon}} \, \lambda^{2p+q} 
%\end{equation}
%
\begin{equation}\label{eq:integrand_expansion_outer}
    \sum_{p , q=0}^\infty \binom{\epsilon/2}{p,q,\epsilon/2-p-q}  (-1)^q \, \frac{x^{2p+q} \, (z+\bar{z})^q}{\abs{z}^{2p+2q-\epsilon}} \, \lambda^{2p+q} 
\end{equation}
can be integrated term by term, yielding
%
%\begin{equation}\label{eq:amplitude_expansion_outer}
%    \A_{xz}^{\overline D} = \sum_{p , q=0}^\infty \frac{(-1)^q \, \Gamma(1+\frac{\epsilon}{2})}{\Gamma(1+\frac{\epsilon}{2}-p-q)} \, \A_x(\alpha,\beta+2p+q) \, \A_z^{\overline D}(a,b-2p-2q+\epsilon,c,d+q) \, \frac{\lambda^{2p+q}}{p! \, q!}\, .
%\end{equation}
%
\begin{equation}\label{eq:amplitude_expansion_outer}
    \A_{xz}^{\overline D} = \sum_{p , q=0}^\infty \binom{\epsilon/2}{p,q,\epsilon/2-p-q} \, \A_x(\alpha,\beta+2p+q) \, \A_z^{\overline D}(a,b-2p-2q+\epsilon,c,d+q) \, (-\lambda)^{2p+q}\, ,
\end{equation}
and the factorized integrals $\A_z^{\overline D}$ that appear in the coefficients can be evaluated using eq.~\eqref{eq:z_integral_complement_solution}.

While the expansions in eqs.~\eqref{eq:amplitude_expansion_inner} and~\eqref{eq:amplitude_expansion_outer} converge in the open unit disk, the integral representations of $\A_{xz}^D(\lambda)$ and $\A_{xz}^{\overline D}(\lambda)$ are continuous~\footnote{Strictly speaking, continuity along the segment $[0,1]$ is sufficient.} at $\lambda = 1$, as we shall discuss in the following section. Thus, one can recover the original amplitude with arbitrary accuracy, either extrapolating the (possibly slowly) convergent series near $\lambda = 1$ or via acceleration methods~\footnote{Further numerical accuracy, with a finite number of coefficients, can be attained via a number of techniques, including Padé approximants and Borel summation~\cite{bender1999advanced}.}.

\subsection{Continuity of the Amplitude}\label{sec:continuity}

We can now show that the integral of eq.~\eqref{eq:total_amplitude}, defined as a holomorphic function of $\lambda$ in the open unit disk via the mixing integrands of eqs.~\eqref{eq:lambda_insertion_disk} and ~\eqref{eq:lambda_insertion_outer} is continuous in the closed segment $0\leq \lambda\leq 1$. This property is instrumental to the evaluation of the amplitude in the case of interest, $\lambda = 1$, since the Taylor series of eqs.~\eqref{eq:amplitude_expansion_inner} and ~\eqref{eq:amplitude_expansion_outer} only converge in the open disk. For $\lambda \in [0,1]$, letting
\begin{eqaed}\label{eq:mixing_f_def}
     f_{D}(\lambda \mid x,z) & \equiv \left[1-\lambda \left(1-\frac{\abs{x-z}^2}{k}\right) \right]^{\epsilon/2} \, , \\
     f_{\overline D}(\lambda \mid x,z) & \equiv \left( 1 + \frac{x^2}{\abs{z}^2} \, \lambda^2 - \frac{x(z+\bar{z})}{\abs{z}^2} \, \lambda \right)^{\epsilon/2} \, ,
\end{eqaed}
for $\Re(\epsilon) \geq 0$ the integral can be uniformly bounded, since 
\begin{eqaed}\label{eq:positive_re_eps_bound}
    \abs{f_{D} (\lambda \mid x,z)} & \leq 1\,,\\ % \left(2 + \frac{\abs{x-z}^2}{k} \right)^{\frac{\Re(\epsilon)}{2}} \, , \\
    \abs{f_{\overline{D}} (\lambda \mid x,z)} & \leq \left( 1 + \frac{x^2}{\abs{z}^2} + \frac{x \, \abs{z+\bar{z}}}{\abs{z}^2} \right)^{\frac{\Re(\epsilon)}{2}} \, ,
\end{eqaed}
and we shall henceforth assume that $\Re(\epsilon) < 0$. In this case,
\begin{eqaed}\label{eq:negative_re_eps_bound}
    \abs{f_{D} (\lambda \mid x,z)} & \leq  \frac{\abs{x-z}^{\Re(\epsilon)}}{k^{\frac{\Re(\epsilon)}{2}}} \, , \\
    \abs{f_{\overline{D}} (\lambda \mid x,z)} & \leq \left(1-\frac{x}{\abs{z}}\right)^{\Re(\epsilon)} \, ,
\end{eqaed}
which feature integrable singularities, since we work within the convergence domain of $\A_{xz}(1)$. These bounds are uniform in $\lambda$ within the interval $[0,1]$, so that one can then take the limit $\lambda \to 1$ under the integral sign by the Lebesgue dominated convergence theorem. Therefore,
\begin{equation}\label{eq:amplitude_limit}
    \A_{xz}(1) = \lim_{\lambda \rightarrow 1^-} \A_{xz}(\lambda) \, ,
\end{equation}
and $\A_{xz}(\lambda)$ is given by the convergent series expansion described in the preceding sections for $0 \leq \lambda < 1$.

\section{Conclusions}\label{sec:conclusions}

Let us briefly summarize our results. In this paper we obtained a class of all-order $\alpha'$ corrections to the effective action of D-brane-anti D-brane systems. These corrections are determined by an expansions in powers of $\alpha'$ of the scattering amplitude for a process involving a closed-string RR state, two open-string tachyons, a world-volume scalar and a world-volume vector that is computed in Section~\ref{sec:amplitude}. In addition, we provided a convergent expansion for this amplitude, thus going beyond the factorized limit that was studied in~\cite{Hatefi:2018qlm} and allowing quantitative insights into the regime that sets in beyond it. This technique is also applicable in other contexts involving integrations of six-point correlators, and we expect that the methods used to derive it may also be relevant to study more complicated cases, including corrections to fermionic couplings~\footnote{For a discussion of all-order fermionic couplings for flux backgrounds in the presence of $\overline{\text{D}3}$-branes, see~\cite{Dasgupta:2016prs}.}.

As we have stressed, unveiling new WZ-like couplings in the effective action of brane systems can be relevant to a number of applications, ranging from (top-down) holographic models~\footnote{For an example where stringy effects in D-brane dynamics play an important role in holography, see~\cite{Schwarz:2013wra}.} to cosmology. All-order $\alpha'$ corrections could grant a firmer quantitative control on vacuum stability in the presence of anti D-branes. This issue which has been thoroughly discussed in the literature, and specifically new couplings, and as we have stressed their all-order corrections, could shed some light on KKLT-like constructions of de Sitter metastable vacua. Finally, the inclusion of all-order $\alpha'$ effects would extend top-down holographic models, such as the Sakai-Sugimoto model of holographic QCD, beyond the (super)gravity approximation, thus extending it to weak values of the 't Hooft coupling.

This type of explicit analysis, aside from its direct bearing on the construction of low-energy effective actions, has the potential to provide instructive insights on curvature corrections, and therefore vexing issues related to supersymmetry breaking, the resulting (in)stabilities~\cite{Basile:2018irz} and the Weak Gravity Conjecture (and extensions thereof)~\cite{ArkaniHamed:2006dz,Ooguri:2016pdq,Obied:2018sgi}. Work in this spirit along the lines of~\cite{Cheung:2018cwt} is in progress.

\section*{Acknowledgements}

Part of this work was completed at the Mathematical Institute of Charles University, at the Wigner Institute and at IHES, ICTP and CERN, which EH would like to thank for their hospitality. IB would like to thank F.~Bascone for useful correspondence. We are also grateful to B.~Jurco, G.~Veneziano, W.~Siegel and P.~Sundell for fruitful discussions. We are especially thankful to A.~Sagnotti for his assistance and feedback on the manuscript.

The authors were supported in part by Scuola Normale Superiore and by INFN (IS CSN4-GSS-PI).

% ---------------------------------------------------------------------------------------------------------------------------------
%		Appendix
% ---------------------------------------------------------------------------------------------------------------------------------

\appendix

\section{Notation for the Factorized Amplitude}\label{appendix:qandns}

In this section we provide explicit expressions for the $N$ and $Q$ functions that appear in eqs.~\eqref{eq:A1},~\eqref{eq:A2},~\eqref{eq:A3} and~\eqref{eq:A4}.
\begin{eqaed}\label{eq:N_functions}
    N_1  & = 2^{-2(t+s+u+v+r+w)-1} \, \frac{\Gamma(-2u+\frac{1}{2}) \, \Gamma(-2t+1)}{\Gamma(-2t-2u+\frac{3}{2})} \, , \nonumber \\
    N_2  & = \frac{\Gamma(-u-r-w) \, \Gamma(-t-v-r+\frac{1}{2}) \, \Gamma(r-s) \, \Gamma(-t-s-u-v-r-w)}{\Gamma(-u-s-w) \, \Gamma(-t-s-v+\frac{1}{2}) \, \Gamma(-u-w-t-v-2r-\frac{1}{2})} \, , \nonumber \\
    N_3 & = 2^{-2(t+s+u+v+r+w)-2} \, \frac{\Gamma(-2u-\frac{1}{2}) \, \Gamma(-2t+1)}{\Gamma(-2t-2u+\frac{1}{2})} \, , \nonumber \\
    N_4 & = \frac{\Gamma(r-s-\frac{1}{2}) \, \Gamma(-r-t-v) \, \Gamma(-r-u-w-\frac{1}{2}) \, \Gamma(-r-s-t-u-v-w-\frac{1}{2})}{ \Gamma(-s-t-v+\frac{1}{2}) \, \Gamma (-s-u-w) \, \Gamma(-2 r-t-u-v-w+\frac{1}{2})} \, , \nonumber \\
    N_5 & = 2^{-2(t+s+u+v+r+w)-2} \, \frac{\Gamma(-2u+\frac{1}{2}) \, \Gamma(-2t)}{\Gamma(-2t-2u+\frac{1}{2})} \, , \nonumber \\
    N_6 & = \frac{\Gamma(r-s-\frac{1}{2}) \, \Gamma(-r-t-v) \, \Gamma(-r-u-w+\frac{1}{2}) \, \Gamma(-r-s-t-u-v-w-\frac{1}{2})}{ \Gamma(-s-t-v+\frac{1}{2}) \, \Gamma(-s-u-w) \, \Gamma(-2 r-t-u-v-w+\frac{1}{2})} \, , \nonumber \\  
    N_7 & = 2^{-2(t+s+u+v+r+w)-3} \, \frac{\Gamma(-2u+\frac{1}{2}) \, \Gamma(-2t)}{\Gamma(-2t-2u+\frac{1}{2})} \, , \nonumber \\
    N_8 & = \frac{\Gamma(r-s) \, \Gamma(-r-t-v-\frac{1}{2}) \, \Gamma(-r-u-w) \, \Gamma(-r-s-t-u-v-w-1)}{\Gamma(-s-t-v-\frac{1}{2}) \, \Gamma(-s-u-w) \, \Gamma(-2 r-t-u-v-w+\frac{1}{2})} \, . \nonumber
\end{eqaed}
\begin{eqaed}\label{eq:Q_functions}
    Q_1 & = \frac{(-r-t-v)}{2(4 t+4 u-1)} \left(r (8 u+2)+s (8 t-4)+8 (t+u) (u+w)+8 t+2 u-2 w-3 \right) \, , \nonumber \\
    Q_2 & = \frac{(r-s-\frac{1}{2}) \, (-2u-\frac{1}{2})}{2 \, (4 t+4 u-3) \, (-2t-2u+\frac{1}{2})} \nonumber \\
    & \quad{} \times \left(r (8 t-8 u-2)+8 t (t+v)-4 t-8 u (u+w)-2 u-4 v+2 w+1 \right) \, , \nonumber \\
    Q_3 & = \frac{(r-s-\frac{1}{2}) \, (-2u-\frac{1}{2}) \, (-r-t-v)}{(-2t-2u+\frac{1}{2})} \, , \nonumber \\
    Q_4 & = \frac{2t (4 r+4 t+4 u-1)+s (8 u-2)+8 v (t+u)+4 u-2 v-1}{2 (4 t+4 u-1)} \, , \nonumber \\
    Q_5 & = \frac{(-2t) \, (-r-t-v)}{(-2t-2u+\frac{1}{2})} \, , \nonumber \\
    Q_6 & = \frac{(-2t) \, (r-s-\frac{1}{2})}{(-2t-2u+\frac{1}{2})} \, , \nonumber \\
    Q_7 & = \bigg(-\frac{4 t \left(r (-2 t+2 u-2 v+2 w+1)+2 s (t-u+v-w)+s+2 t+2 v+1 \right)}{(4 t+4 u-1) \, (2 s+2 t+2 v+1)}
    \nonumber \\
    & \quad{} +r+t+v+\frac{1}{2}\bigg) \!\left(-w-\frac{1}{2} \right) \, , \nonumber \\
    Q_8 & = \frac{ (-s-\frac{1}{4}) \, (-2t) \, (-2 r-t-u-v-w-\frac{1}{2})}{(-2t-2u+\frac{1}{2})} \nonumber \\
    & \quad{} \times \bigg(-\frac{2 \left(4 r^2+r (-4 s+4 u+4 w+2)-4 s (u+w)+2 t+2 v+1\right)}{(2 s+2 t+2 v+1)  \, (4 r+2 t+2 u+2 v+2 w+1)} \nonumber \\
    & \quad{} +\frac{2 \, (2 t-1) \, (2 r+2 t+2 v+1)}{(4 t+4 u-3) \, (2 s+2 t+2 v+1)}-\frac{2 \, (r+u+w)}{4 r+2 t+2 u+2 v+2 w+1}\bigg)   \,,\nonumber\\
    Q_9 & = \frac{ (-v-\frac{1}{4}) \, (-2t)}{2 \, (-2u-\frac{1}{2}) \, (4 t+4 u-1) \, (-s-t-v-\frac{1}{2})} 
    \nonumber \\ 
    & \quad{} \times \bigg(2 \, (r-s) \, \big(r (8 t+8 u-2)-4 \left(t^2+t (-3 u+v-3 w)-2 u (u+w)\right) \\
    & \quad{}-4 u+2 v-4 w+1 \big)+(4 t+4 u-1) \, (r+u+w) \, (2 s+2 t+2 v+1) \\
    & \quad{} +(8 t+4 u-3) \, (2 r+2 t+2 v+1) \bigg) \, . \nonumber
\end{eqaed}

\section{Computation of $\A_z$, $\A_z^D$ and $\A_z^{\overline D}$}\label{appendix:z_integral}

In this section we derive a closed-form expression for the integral in eq.~\eqref{eq:z_amplitude},
\begin{equation}\label{eq:z_amplitude_appendix}
    \A_z \equiv \A_z(a,b,c,d) = \int_\Hp d^2 z \, \abs{1-z}^a \abs{z}^b \,(z-\bar{z})^c (z+\bar{z})^d \, .
\end{equation}
We restrict ourselves to $d \in \mathbb{N}$, which is the case of interest. To begin with, one can write
\begin{eqaed}\label{eq:exp_trick}
    \abs{z}^b & = \frac{1}{\Gamma(-\frac{b}{2})} \int_0^\infty du\, u^{-\frac{b}{2}-1} e^{-\abs{z}^2 u} \, ,   \\
    \abs{1-z}^a & = \frac{1}{\Gamma(-\frac{a}{2})} \int_0^\infty dt\, t^{-\frac{a}{2}-1} e^{-\abs{1-z}^2 t} \, .
\end{eqaed}
Letting $z=x+iy$, 
\begin{equation}\label{eq:z_integral}
    \A_z = \frac{(2i)^{c}\, 2^d \,(2i)}{\Gamma(-\frac{a}{2})\Gamma(-\frac{b}{2})} \int_0^\infty dy \int_{-\infty}^\infty dx \, y^c x^d \int_0^\infty du \int_0^\infty dt \, u^{-\frac{b}{2}-1} t^{-\frac{a}{2}-1} e^{-(\abs{z}^2 u + \abs{1-z}^2 t)} \, ,
\end{equation}
%
%and completing the square in the exponential,
%
%\begin{equation}\label{eq:rewriteexp}
%    e^{-(\abs{z}^2 u + \abs{1-z}^2 t)} = e^{-(x^2(u+t) - 2xt \, + y^2(u+t) \, +t)} \, ,
%\end{equation}
%
%reduces the problem to known integrals. Indeed, the integral over $y$ reads
%
%\begin{equation}\label{eq:y_integral}
%    \int_0^\infty dy \, y^c \, e^{-y^2(u+t)}
%\end{equation}
%
%which, upon the substitution $v = y^2(u+t)$, $dy=\frac{dv}{2\sqrt{(u+t)v}}$ becomes a Gamma function,
%
%\begin{equation}\label{eq:y_gamma}
%   \frac{1}{2(u+t)^{\frac{1+c}{2}} } \int_0^\infty dv \, v^{\frac{c-1}{2}} \, e^{-v} = \frac{\Gamma(\frac{1+c}{2})}{2(u+t)^{\frac{1+c}{2}}} \, .
%\end{equation}
%
%Then, selecting only the terms depending on $x$, the integral over $x$ is given by
%
%\begin{equation}\label{eq:x_integral}
%    \int_{-\infty}^\infty dx \,x^d e^{-((u+t)x^2-2tx)} \, .
%\end{equation} 
%
%Substituting $r = \sqrt{u+t} \,x$ yields
%
%\begin{equation}\label{eq:x_preresult}
%    \frac{1}{(u+t)^{\frac{1+d}{2}}} \int_{-\infty}^\infty dr \, r^d \, e^{-r^2+\frac{2t}{\sqrt{u+t}} r} = \frac{1}{2^d (u+t)^{\frac{1+d}{2}}}  \dv[d]{F(\alpha)}{v}\Big|_{v=\frac{t}{\sqrt{u+t}}} \, ,
%\end{equation}
%
%where
%
%\begin{equation}\label{eq:Fgen_def}
%    F(v) \equiv \int_{-\infty}^\infty dr \,e^{-r^2+2v r} = \sqrt{\pi} \, e^{v^2} \, .
%\end{equation}
%
so that the integral over $y$ gives
\begin{equation}\label{eq:y_integral}
    \int_0^\infty dy \, y^c \, e^{-y^2(u+t)} = \frac{\Gamma(\frac{1+c}{2})}{2(u+t)^{\frac{1+c}{2}}} \, ,
\end{equation}
while, in terms of the polynomials defined in eq.~\eqref{eq:polydef}, the integral over $x$ reduces to
\begin{equation}\label{eq:x_integral}
    \int_{-\infty}^\infty dx \,x^d e^{-((u+t)x^2-2tx)} = \frac{\sqrt{\pi}}{(u+t)^{\frac{1+d}{2}}2^d} \,  P_d\left(\frac{t}{\sqrt{u+t}}\right) \, e^\frac{t^2}{u+t} \, .
\end{equation} 
All in all,
\begin{eqaed}\label{eq:z_preresult}
    \A_z & = \frac{(2i)^{c+1} \sqrt{\pi} \, \Gamma(\frac{1+c}{2}) }{\Gamma(-\frac{a}{2}) \Gamma(-\frac{b}{2})} \int_0^\infty du \int_0^\infty dt \, u^{-\frac{b}{2}-1} \,  t^{-\frac{a}{2}-1} \, e^{-\frac{ut}{u+t}}   \\
    & \times (u+t)^{-1-\frac{(c+d)}{2}} \, P_d\left(\frac{t}{\sqrt{u+t}}\right) \, ,
\end{eqaed}
and each monomial in $P_d$ gives rise to an integral of the form
\begin{equation}\label{eq:I_k}
    I_k = \int_0^\infty du \int_0^\infty dt\, u^{-\frac{b}{2}-1} t^{-\frac{a}{2}-1} e^{-\frac{ut}{u+t}} t^k (u+t)^{-1-\frac{c+d+k}{2}} \, .
\end{equation}
%
%which can be evaluates via the substitutions
%
%\begin{equation}\label{eq:t_u_vars}
%    t=\frac{x}{s}\,,\qquad u = \frac{x}{1-s}\,,\qquad dt \wedge du = \frac{x}{(s(1-s))^2} \, dx \wedge ds \, .
%\end{equation}
%
After the substitution
\begin{equation}\label{eq:t_u_vars}
    t=\frac{x}{s}\,,\qquad u = \frac{x}{1-s}\,,\qquad dt \wedge du = \frac{x}{(s(1-s))^2} \, dx \wedge ds
\end{equation}
one finds
\begin{eqaed}\label{eq:I_k_result}
    I_k = \Gamma\left(-1-\frac{a+b+c+d-k}{2}\right) \Beta\left(1+\frac{a+c+d-k}{2},1+ \frac{b+c+d+k}{2}\right)\,,
\end{eqaed}
so that the final result reads
\begin{eqaed}\label{eq:z_result}
    \A_z & = \frac{(2i)^{1+c} \sqrt{\pi}}{2} \frac{\Gamma(\frac{1+c}{2})}{\Gamma(-\frac{a}{2})\Gamma(-\frac{b}{2}) } \sum_{k=0}^d h_{k,d} \, \Gamma\left(-1-\frac{a+b+c+d-k}{2}\right)   \\
    & \times \Beta\left(1+\frac{a+c+d-k}{2}, 1+\frac{b+c+d+k}{2}\right) \, .
\end{eqaed}

Let us now discuss the analogous integral on the upper-half unit disk,
\begin{equation}\label{eq:z_integral_disk}
\A^D_z(a,b,c,d) = \int_{\Hp, \, \abs{z}<1} \!\!\!\!\!\!\! d^2 z \, \abs{1-z}^a \abs{z}^b \,(z-\bar{z})^c (z+\bar{z})^d \, ,
\end{equation}
recasting it as a one-dimensional integral suitable for numerical evaluation. To this end, we change variables according to $z = r(t+i\sqrt{1-t^2})$, with $0<r<1$, $-1<t<1$, so that
\begin{eqaed}\label{eq:z_integral_disk_polar}
    \A^D_z & = 2^{d} \, (2i)^{1+c} \int_0^1 dr \, r^{1+b+c+d} \, (1+r^2)^{a/2} \\
    & \quad{} \times \int_{-1}^1 dt \, \left( 1- \frac{2r}{1+r^2} \, t\right)^{a/2} t^d \, (1-t)^{\frac{c-1}{2}} (1+t)^{\frac{c-1}{2}}\,.
\end{eqaed}
Recalling the integral representation
\begin{equation}\label{eq:F1_appell}
    \int_0^1 dx \, x^{\lambda-1} (1-x)^{\mu-1} (1-u \, x)^{-\rho} (1-v \, x)^{-\sigma} \, dx = \Beta(\mu,\lambda) \, F_1(\lambda,\rho,\sigma,\lambda+\mu;u,v)
\end{equation}
of the Appell series $F_1$~\cite{bailey1935generalized}, one can rewrite the integral as
\begin{eqaed}\label{eq:z_integral_disk_solution}
    \A_z^D(a,b,c,d) & = 2^{d} (2i)^{1+c} \, \Beta \left(\frac{c+1}{2},d+1\right)\\
    &\quad{}\times\int_0^1 dr \,r^{1+b+c+d}(1+r^2)^{a/2} \left(F_{abcd}(r) + (-1)^d F_{abcd}(-r)\right) \, ,
\end{eqaed}
where we have defined
\begin{eqaed}\label{eq:F1_contracted}
     F_{abcd}(r) & \equiv F_1\left(d+1,-\frac{a}{2},\frac{1-c}{2}, d+\frac{c}{2}+\frac{3}{2}; \frac{2r}{1+r^2},-1\right) \\
     & = 2^{-(d+1)} F_1\left(d+1, \, -\frac{a}{2},\, \frac{a}{2}+c+d+1;\, \frac{(1+r)^2}{2(1+r^2)}, \frac{1}{2} \right)\,.%\label{eq:F1_flipped}
\end{eqaed}
The second line follows from eq.~(9.4.3) in~\cite{bailey1935generalized}, and is better suited for numerical applications, since the Appell series is evaluated within its domain of convergence.
%where the equality follows from Eq. 9.4.3 in~\cite{bailey1935generalized}. Form \eqref{eq:F1_flipped} is preferrable to \eqref{eq:F1_contracted} for numerical applications because the Appell series is evaluated within the domain of convergence.

Let us finally consider the integral $\A_z^{\overline D}$, whose domain is the complement of the unit disk in $\Hp$,
\begin{equation}\label{eq:z_integral_complement}
\A^{\overline{D}}_z(a,b,c,d) = \int_{\Hp, \, \abs{z}>1} \!\!\!\!\!\!\! d^2 z \, \abs{1-z}^a \abs{z}^b \,(z-\bar{z})^c (z+\bar{z})^d \, .
\end{equation}
Changing variables according to $z \rightarrow 1/\bar{z}$ recasts the integral in the form
\begin{equation}\label{eq:z_integral_complement_reciprocal}
    \A^{\overline D}_z(a,b,c,d) = \int_{\Hp, \, \abs{z}<1} \!\!\!\!\!\!\! d^2 z \, \abs{z}^{-2-a-b-2c-2d} \, \abs{1-z}^a (z-\bar{z})^c (z+\bar z)^d \, ,
\end{equation}
so that
\begin{equation}\label{eq:z_integral_complement_solution}
    \A^{\overline{D}}_z(a,b,c,d) = \A_z^D(a,-2-a-b-2c-2d,c,d) \, .
\end{equation}
Hence, one is led back to the integral representation of eq.~\eqref{eq:z_integral_disk_solution}.

\subsection{The $P_n$ Polynomials}\label{appendix:polynomials}

We describe now the polynomials associated to the coefficients $h_{k,n}$, which are needed in eq.~\eqref{eq:z_result}.%, and we briefly discuss their large-order asymptotics used in Section~\ref{sec:borel}.

Let us define the family polynomials
\begin{equation}\label{eq:polydef}
    P_n(v) \equiv e^{-v^2} \,  \dv[n]{e^{v^2}}{v} \, ,
\end{equation}
which are related to the physicists' Hermite polynomials via $H_n(iv) = i^n P_n(v)$. One can also define them via the recurrence relation
\begin{equation}\label{eq:recurrence}
    P_n(v) = \left(2v + \dv{}{v}\right) P_{n-1}(v) \, , \qquad P_0(v) = 1 \, ,
\end{equation}
and the corresponding coefficients,
\begin{equation}\label{eq:P_coefficients}
    P_n(v) \equiv \sum_{k=0}^n h_{k,n} \, v^k \, ,
\end{equation}
are manifestly \textit{positive}. These coefficients enter the closed-form expression for the integral in eq.~\eqref{eq:z_amplitude}.

\end{document}